\documentclass[aps,pra,twocolumn,10pt,notitlepage,showpacs]{revtex4-1}

\usepackage{tikz}
\usetikzlibrary{calc,patterns}
\usepackage{ifthen}
\usepackage{amssymb}
\usepackage{mathtools}
\usepackage{dsfont}
\usepackage{bbm}
\usepackage{mleftright}\mleftright
\usepackage{thmtools} 
\usepackage{thm-restate}
\usepackage{color}
\usepackage{xcolor}

\usepackage{algorithm2e}
\usepackage{enumitem}
\usepackage{float}

\usepackage{caption}
\usepackage{subcaption}

\newcommand{\eps}{\varepsilon}

\newcommand{\braketbra}[3]{\langle #1|#2| #3 \rangle}
\newcommand{\nrm}[1]{\left\lVert #1 \right\rVert}
\newcommand{\bigO}[1]{\mathcal{O}\left( #1 \right)}

\newcommand{\pvp}{\vec{p}{\kern 0.45mm}'}
\let\oldnabla\nabla
\renewcommand{\nabla}{\oldnabla\!}

\DeclarePairedDelimiter\bra{\langle}{\rvert}
\DeclarePairedDelimiter\ket{\lvert}{\rangle}
\DeclarePairedDelimiterX\braket[2]{\langle}{\rangle}{#1 \delimsize\vert #2}
\newcommand{\underflow}[2]{\underset{\kern-60mm \overbrace{#1} \kern-60mm}{#2}}

\long\def\ignore#1{}

\newtheorem{theorem}{Theorem}

\newtheorem{lemma}[theorem]{Lemma}
\newtheorem{fact}[theorem]{Fact}

\newtheorem{definition}[theorem]{Definition}

\newcommand{\G}{\ensuremath{\mathcal{G}}}

\newcommand{\C}{\ensuremath{\mathcal{C}}}

\newcommand{\Oo}{\ensuremath{\mathcal{O}}}

\newcommand{\R}{\ensuremath{\mathbb{R}}}

\newcommand{\U}{\ensuremath{\mathcal{U}}}

\newcommand{\E}{\mathcal{E}}
\newcommand{\M}{\mathcal{M}}
\usepackage{tikz}

\usetikzlibrary{backgrounds,fit,decorations.pathreplacing,calc}

\usepackage{qcircuit}

\newenvironment{proof}
{\noindent {\bf Proof. }}
{{\hfill $\Box$}\\	\smallskip}

\usepackage[final]{hyperref}
\hypersetup{
	colorlinks = true,
	allcolors = {blue},
}

\usepackage{comment}
\usepackage{amsmath}
\usepackage{graphicx}
\usepackage{subcaption}
\usepackage{caption}
\usepackage{amssymb}
\usepackage{thm-restate}
\usepackage{amsfonts}
\usepackage[parfill]{parskip}
\usepackage{hyperref}

\usepackage[format=plain,justification=centerlast,singlelinecheck=true]{caption}

\begin{document}


\title{Finding a marked node on any graph by continuous-time quantum walk}


\author{Shantanav Chakraborty}  
\email[]{shchakra@ulb.ac.be}
\author{Leonardo Novo}
\email[]{lfgnovo@gmail.com}
\author{J\'{e}r\'{e}mie Roland}
\email[]{jroland@ulb.ac.be}
\affiliation{QuIC, Ecole Polytechnique de Bruxelles, Universit\'{e} Libre de Bruxelles}
\begin{abstract}
Spatial search by discrete-time quantum walk can find a marked node on any ergodic, reversible Markov chain $P$ quadratically faster than its classical counterpart, i.e.\ in a time that is in the square root of the hitting time of $P$. However, in the framework of continuous-time quantum walks, it was previously unknown whether such general speed-up is possible. In fact, in this framework, the widely used quantum algorithm by Childs and Goldstone fails to achieve such a speedup. Furthermore, it is not clear how to apply this algorithm for searching any Markov chain $P$. In this article, we aim to reconcile the apparent differences between the running times of spatial search algorithms in these two frameworks. We first present a modified version of the Childs and Goldstone algorithm which can search for a marked element for any ergodic, reversible $P$ by performing a quantum walk on its edges. Although this approach improves the algorithmic running time for several instances, it cannot provide a generic quadratic speedup for any $P$. Secondly, using the framework of interpolated Markov chains, we provide a new spatial search algorithm by continuous-time quantum walk which can find a marked node on any $P$ in the square root of the classical hitting time. In the scenario where multiple nodes are marked, the algorithmic running time scales as the square root of a quantity known as the extended hitting time. Our results establish a novel connection between discrete-time and continuous-time quantum walks and can be used to develop a number of Markov chain-based quantum algorithms.
\end{abstract}
\date{\today}
\maketitle

\section{Introduction}
The problem of finding a set of marked nodes in a graph, known as the spatial search problem, can be tackled using a random walk. The expected number of steps required by the walker to find a node within this marked set is known as the hitting time of the random walk. Quantum walks, which are quantum analogues of classical random walks, also provide a natural framework to tackle this problem. For discrete-time quantum walks (DTQW), it has been established that the time required to find a single marked node on any ergodic, reversible Markov chain is quadratically faster than its classical counterpart \cite{krovi2016quantum}. 

However, the situation is drastically different for continuous-time quantum walks (CTQW) where quadratic speed-ups were known to be possible only for specific examples. The first spatial search algorithm by CTQW was introduced by Childs and Goldstone \cite{childs2004spatial} (which we shall refer to as the $\C\G$ algorithm). Therein, the authors demonstrated that a marked node among $n$ nodes can be found in $\mathcal{O}(\sqrt{n})$ time for certain graphs such as the complete graph, hybercube and $d$-dimensional lattices with $d>4$,  implying a quadratic speedup for the spatial search problem with respect to classical random walks for these graphs. However when $d=4$, the running time of the $\C\G$ algorithm is in $\mathcal{O}(\sqrt{n}\log n)$ and so a full quadratic speedup is lost. In fact, there exists no substantial speedup for lattices of dimension less than four. Since then a plethora of results have been published exhibiting a $\mathcal{O}(\sqrt{n})$ running time of the $\C\G$ algorithm on certain specific graphs~\cite{janmark2014global, meyer2015connectivity, novo2015systematic, philipp2016continuous, wong2016quantum} or ensembles of random graphs \cite{chakraborty2016spatial, chakraborty2017optimal}. Recently, we obtained the necessary and sufficient conditions for the $\C\G$ algorithm to be optimal for any graph that meets certain general spectral properties \cite{shanleonew2019}. From this result, one can recover most of the prior results on the optimality of this algorithm for specific graphs. Furthermore, by exploring the regimes where these optimality conditions are violated,  we have also provided new examples of graphs for which this algorithm fails to attain a generic quadratic speedup \cite{shanleonew2019}. 

To summarize, the original $\C\G$ formalism for spatial search by CTQW suffers from the following limitations:  it does not provide general quadratic speed-ups for the problem of finding a single marked node on a graph; general results on its performance when multiple nodes are marked are not available; it is unclear how to apply this algorithm to any ergodic, reversible Markov chain. These limitations imply that the current state of the art for spatial search by CTQW is far from matching its discrete-time counterpart. 

The aim of this article is to reconcile this apparent gap between CTQW and DTQW for the spatial search problem. Our main contributions are two new algorithms for spatial search by CTQW. The first algorithm can be seen as a modified version of the Childs and Goldstone formalism, involving a CTQW on the edges of any ergodic, reversible Markov chain. We show that this algorithm improves the performance of the original Childs and Goldstone algorithm on several important instances, but a general quadratic speedup remains elusive. Our second algorithm goes beyond the Childs and Golstone formalism and performs a CTQW on the edges of an \emph{interpolated Markov chain}, inspired by the adiabatic algorithm for finding marked nodes on a Markov chain by adiabatic evolution ~\cite{krovi2010adiabatic}. This new CTQW-based spatial search algorithm can find a marked node on any ergodic, reversible Markov chain in square root of the hitting time of the corresponding classical random walk. Moreover, it has a guaranteed performance in the scenario where multiple nodes are marked, finding a marked node in square root of a quantity known as the \emph{extended hitting time} \cite{krovi2010adiabatic, krovi2016quantum}. 

Our techniques are also inspired by the progress made in the DTQW framework for the spatial search problem. While the $\C\G$ algorithm is defined on a Hilbert space of dimension $n$,  where $n$ is the number of nodes of the graph, it was observed by Ambainis, Kempe and Rivosh that additional (coin) degrees of freedom, storing the direction in which the walker moves, can make DTQW faster. In fact, they demonstrated that the running time of the spatial search algorithm by coined DTQW is in $\Oo(\sqrt{n}\log n)$ even for $2d$ lattices \cite{ambainis2005coins} and $\Oo(\sqrt{n})$ for $d>2$, thereby outperforming the original $\C\G$ algorithm. It is thus natural to ask whether additional degrees of freedom can make CTQW faster. The result of Ref.~\cite{childs2004spatial-dirac} seems to suggest so for $2d$-lattices: the introduction of a spin degree of freedom has helped improve the running time of the $\C\G$ algorithm to $\mathcal{O}(\sqrt{n}\log n)$. 

Furthermore, in DTQW framework, the spatial search algorithm can be applied to Markov chains. Inspired by Ambainis' algorithm for element distinctness \cite{ambainis2007quantum}, Szegedy provided a general technique to construct a quantum analogue of any ergodic, reversible Markov chain \cite{szegedy2004quantum}. The crucial difference of Szegedy's quantum walk from prior works is that the underlying walk takes place on an enlarged state-space, namely on the edges of the Markov chain, instead of its vertices. The resulting increase in the Hilbert space dimension acts as a generalized \textit{coin}. Szegedy's work resulted in subsequent works on DTQW-based spatial search algorithms \cite{santha2008quantum, magniez2011search}, leading to the work of Krovi et al.~\cite{krovi2016quantum}. The algorithm in Ref.~\cite{krovi2016quantum} finds a node in a marked set of vertices on any ergodic, reversible Markov chain in a time that is the square root of a quantity known as the \textit{extended hitting time}. As \textit{extended hitting time} is the same as \textit{hitting time} for a single marked node, this implied a full quadratic speedup in this scenario. Naturally, one can ask whether CTQW-based spatial search algorithms can be applied more generally to ergodic, reversible Markov chains where multiple nodes are marked. If so, can one expect that CTQW on the edges of the underlying Markov chain would yield a general quadratic speedup as is the case of the DTQW?

In this article, we make significant progress towards answering these questions. Our contributions are summarized below:
\\~\\
\textbf{\textit{(i)~Modified $\C\G$ algorithm ($\C\G'$):}~} As mentioned previously, in the framework of DTQW, it has been established that external degrees of freedom (coins) improve the running time of the spatial search algorithm \cite{ambainis2005coins} and that for general graphs this represents a walk on the edges of the underlying graph instead of nodes \cite{szegedy2004quantum}. Inspired by this fact, in Sec.~\ref{sec-main:modified-cg-algo} we show that the $\C\G$ algorithm can be modified to yield a CTQW-based spatial search algorithm for any ergodic, reversible Markov chain $P$. This new algorithm (which we shall refer to as the $\C\G'$ algorithm) uses the Hamiltonian constructed by the formalism of Somma and Ortiz \cite{somma2010quantum} to encode $P$, in addition with an oracular Hamiltonian of a similar form of the one defined in Refs.~\cite{foulger2014quantum, childs2014spatial-crystal}. We show that the Hamiltonian by Somma and Ortiz can be seen as a quantum walk on the edges of $P$. Interestingly, the $\C\G'$ algorithm improves upon the running time of the $\C\G$ algorithm for several instances; for example, when applied to $2d$-lattices, the $\C\G'$ algorithm has a running time of $\Theta(\sqrt{n}\log n)$. However it fails to attain a generic quadratic speedup over classical random walks. Also, as with the $\C\G$ algorithm, it is unclear what is the performance of the algorithm when multiple vertices are marked.
\\~\\
\textbf{\textit{(ii)~CTQW-based spatial search algorithm with a generic quadratic speedup:}~} In Sec.~\ref{sec-main:search-algo-phase-randomization}, we provide a new CTQW-based spatial search algorithm that can find a marked vertex on any ergodic, reversible Markov chain in the square root of the classical hitting time. Our algorithm also has a guaranteed performance in the scenario where there are multiple marked vertices. In fact, for any ergodic, reversible Markov chain $P$ with a set of $M$ marked vertices, it runs in the square root of the \textit{extended hitting time}. Inspired by the techniques of Refs.~\cite{krovi2010adiabatic, krovi2016quantum},  we construct, for a given Markov chain $P$, a quantum analogues of the interpolating Markov Chain, $P(s)=(1-s)P+sP'$, where $P'$ is obtained from $P$ by replacing all outgoing edges from $M$ by self loops. This algorithm involves choosing an appropriate value of $s=s^*$ and evolve a time-independent Hamiltonian encoding $P(s^*)$ for a time that is chosen uniformly at random in the interval $[0,T]$, followed by a  measurement in the basis of the state space of the underlying Markov chain.  We prove that provided $T$ (which is also the expected running time of our algorithm) scales as the square root of the extended hitting time, a vertex from $M$ can be obtained with a high probability. In Ref.~\cite{krovi2010adiabatic}, the same Hamiltonian was used to solve the spatial search problem by adiabatic evolving the parameter $s$, with a similar performance. Here we prove that it is possible to bypass the adiabatic evolution altogether and still obtain the same running time. 
\section{Preliminaries}
\label{sec-main:preliminaries}
In this section, we review some basic concepts about Markov chains and interpolated Markov chains focusing on the main spectral properties that will be needed to analyse our CTQW algorithms. Furthermore, we introduce the notions of \emph{hitting time} and \emph{extended hitting time} of classical random walks.
\subsection{Basics on Markov chains}
A Markov chain on a discrete state space $X$, such that $|X|=n$, can be described by a $n\times n$ stochastic matrix $P$ \cite{norris1998markov}. Each entry $p_{xy}$ of this matrix $P$ represents the probability of transitioning from state $x$ to state $y$. Any distribution over the state space of the Markov chain is represented by a stochastic row vector. 

A Markov chain is \textit{irreducible} if any state can be reached from any other state in a finite number of steps. Any \textit{irreducible} Markov chain is \textit{aperiodic} if there exists no integer greater than one that divides the length of every directed cycle of the graph. A Markov chain is \textit{ergodic} if it is both \textit{irreducible} and \textit{aperiodic}. By the Perron-Frobenius Theorem, any ergodic Markov chain $P$ has a unique stationary state $\pi$ such that $\pi P=\pi$. The stationary state $\pi$ is a stochastic row vector and has support on all the elements of $X$. Let us denote it as 
\begin{equation}
\label{eqmain:stationary-state-classical}
\pi=\left(\pi_1~~\pi_2~~\cdots~~\pi_n\right),
\end{equation}
such that $\sum_{j=1}^n \pi_j=1$. 

Starting from any initial probability distribution $\mu$ over the state space $X$, the repeated application of $P$ leads to convergence to the stationary distribution $\pi$, i.e.\ $\lim_{t\rightarrow\infty} \mu P^t=\pi$. It follows from the Perron-Frobenius theorem that other than $\pi$, all eigenvectors have eigenvalues of absolute value strictly less than $1$. Thus, $\pi$ is the unique eigenvector with eigenvalue $1$ and all other eigenvalues lie between $-1$ and $1$. Throughout the paper we shall be working with the Markov chain corresponding to the \textit{lazy walk}, i.e.\ we shall map $P\mapsto (I+P)/2$. This transformation ensures that all the eigenvalues of $P$ lie between $0$ and $1$ and will not affect our results other than by a factor of two, which is irrelevant for the assymptotic running time of our algorithms. An important quantity throughout this work is the gap between the two highest eigenvalues of $P$ (the spectral gap), which we denote by $\Delta$.

Let $p_{x,y}$ denote the $(x,y)^{\text{th}}$-entry of the ergodic Markov chain $P$ with stationary state $\pi$. Then the $(x,y)^{\text{th}}$ entry of the time-reversed Markov chain of $P$, denoted by $P^*$, is 
\begin{equation}
p^*_{x,y}=p_{y,x}\dfrac{\pi_y}{\pi_x}.    
\end{equation}
We shall concern ourselves with ergodic Markov chains that are also \textit{reversible}, i.e. Markov chains for which $P=P^*$. Any reversible $P$ satisfies the \textit{detailed balance condition} 
\begin{equation}
\pi_x p_{xy}=\pi_y p_{yx},~\forall (x,y)\in X.    
\end{equation} 
This can also be rewritten as 
\begin{equation}
\text{diag}(\pi)P=P^T\text{diag}(\pi),    
\end{equation}
where $\text{diag}(\pi)$ is a diagonal matrix with the $j^{\text{th}}$ diagonal entry being $\pi_j$. In other words, the reversibility criterion implies that the matrix $\text{diag}(\pi)P$ is symmetric. Henceforth we shall only deal with ergodic reversible Markov chains.
~\\
\textbf{Discriminant matrix:~} The discriminant matrix of $P$ is defined as 
\begin{equation}
\label{eqmain:discriminant-matrix-definition}
D(P)=\sqrt{P\circ P^T},
\end{equation}
where $\circ$ indicates the Hadamard product and the $(x,y)^{\mathrm{th}}$ entry of $D(P)$ is $D_{xy}(P)=\sqrt{p_{xy}p_{yx}}$. Thus $D(P)$ is a symmetric matrix.

For any reversible Markov chain $P$, we have that 
\begin{equation}
D(P)=\mathrm{diag}(\sqrt{\pi})P\mathrm{diag}(\sqrt{\pi})^{-1},
\end{equation}
where $\sqrt{\pi}$ is a row vector with its $j^{\text{th}}$-entry being $\sqrt{\pi_j}$. This can be derived from the detailed-balance condition, which allows to express each entry of $D(P)$ as 
\begin{align}
D_{xy}(P)&=\sqrt{p_{xy}p_{yx}}\\
            &=p_{xy}\sqrt{\frac{\pi_x}{\pi_y}}.
\end{align}
From this fact, we obtain that $D(P)$ is \textit{similar} to $P$, i.e.\ they have the same set of eigenvalues. So if the eigenvalues of $P$ are ordered as $\lambda_n=1 > \lambda_{n-1}\geq\cdots\geq \lambda_1$, the spectral decomposition of $D(P)$ is
\begin{equation}
\label{eqmain:discriminant-matrix-spectral}
D(P)=\sum_{i=1}^{n}\lambda_i\ket{v_i}\bra{v_i},
\end{equation}
where $\ket{v_i}$ is an eigenvector of $D(P)$ with eigenvalue $\lambda_i$. 

Note that from the reversibility condition stated previously,
\begin{align}
D(P)\sqrt{\pi^T}&=\mathrm{diag}(\sqrt{\pi})P\mathrm{diag}(\sqrt{\pi})^{-1}\sqrt{\pi^T}\\
                      &=\sqrt{\pi^T}.
\end{align}
Thus, the eigenstate of $D(P)$ with eigenvalue $1$ is given by
\begin{equation}
\label{eqmain:initial-stationary-state}
    \ket{v_n}=\sqrt{\pi^T}=\sum_{x\in X}\sqrt{\pi_x}\ket{x}.
\end{equation}
\subsection{Interpolated Markov chains}\label{sec:interpolatedMC}
Let us assume that a subset of the elements of the state space of the Markov chain $P$ is marked. Let $M\subset X$ denote the set of marked elements. Given any $P$, we define $P'$ as the \textit{absorbing Markov chain} obtained from $P$ by replacing all the outgoing edges from $M$ by self-loops. If we re-arrange the elements of $X$ such that the unmarked elements $U:=X\backslash M$ appear first, then we can write 
\begin{align}\label{eqmain:PandPprime}
P=\begin{bmatrix}
P_{UU} & P_ {UM}\\
P_{MU} & P_{MM}
\end{bmatrix},~~~~~~~~~P'=\begin{bmatrix}
P_{UU} & P_ {UM}\\
0 & I
\end{bmatrix},
\end{align}
where $P_{UU}$ and $P_{MM}$ are square matrices of size $(n-|M|)\times (n-|M|)$ and $|M|\times |M|$ respectively. On the other hand $P_{UM}$ and $P_{MU}$ are matrices of size $(n-|M|)\times |M|$ and $|M|\times (n-|M|)$ respectively. Then the \textit{interpolated Markov chain} is defined as 
\begin{equation}
\label{eqmain:interpolated-mc-defintion}
P(s)=(1-s)P+sP',
\end{equation}
where $s\in[0,1]$. The interpolated Markov chain thus has a block structure
\begin{align}\label{eqmain:Pofs}
P(s)=\begin{bmatrix}
P_{UU} & P_ {UM}\\
(1-s)P_{MU} & (1-s)P_{MM}+sI
\end{bmatrix}.
\end{align}
Clearly, $P(0)=P$ and $P(1)=P'$. Notice that if $P$ is ergodic, so is $P(s)$ for $s\in [0,1)$. This is because any edge in $P$ is also an edge of $P(s)$ and so the properties of \textit{irreducibility} and \textit{aperiodicity} are preserved. However when $s=1$, $P(s)$ has outgoing edges from $M$ replaced by self-loops and as such the states in $U$ are not accessible from $M$, implying that $P(1)$ is not ergodic. We denote the spectral gap of $P(s)$ as $\Delta(s)$. 

Now we shall see how the stationary state of $P$ is related to that of $P(s)$. Since $X=U \cup M$, the stationary state $\pi$ can be written as 
\begin{equation}
\label{eqmain:stationary-state-split}
\pi=(\pi_U~~\pi_M),
\end{equation}
where $\pi_U$ and $\pi_M$ are row-vectors of length $n-|M|$ and $|M|$ respectively. As mentioned previously, $P'$ is not ergodic and does not have a unique stationary state. In fact, any state having support over only the marked set is a stationary state of $P'$. 

On the other hand $P(s)$ is ergodic for $s\in [0,1)$. Let $p_M=\sum_{x\in M}\pi_x$ be the probability of obtaining a marked element in the stationary state of $P$. Then it is easy to verify that the unique stationary state of $P(s)$ is 
\begin{equation}
\label{eqmain:stationary-state-interpolated-mc-classical}
\pi(s)=\dfrac{1}{1-s(1-p_M)}\left((1-s)\pi_U~~\pi_M\right).
\end{equation}
~\\~\\
The discriminant matrix of $P(s)$ is defined as 
\begin{equation}
\label{eqmain:discriminant-matrix-definition}
D(P(s))=\sqrt{P(s)\circ P(s)^T},
\end{equation}
where $\circ$ indicates the Hadamard product and the $(x,y)^{\mathrm{th}}$ entry of $D(P(s))$ is $D_{xy}(P(s))=\sqrt{p_{xy}(s)p_{yx}(s)}$. Thus $D(P(s))$ is a symmetric matrix.
~\\
Let the spectral decomposition of $D(P(s))$ be
\begin{equation}
\label{eqmain:discriminant-matrix-spectral_s}
D(P(s))=\sum_{i=1}^{n}\lambda_i(s)\ket{v_i(s)}\bra{v_i(s)},
\end{equation}
where $\ket{v_i(s)}$ is an eigenvector of $D(P(s))$ with eigenvalue $\lambda_i(s)$, such that $\lambda_n(s)=1 > \lambda_{n-1}(s)\geq\cdots\geq \lambda_1(s)$. 
~\\
It can be seen that the eigenstate of eigenvalue $1$ of $D(P(s))$ can be expressed as 
\begin{align}
\label{eqmain:highest-eigenstate-discriminant-matrix}
\ket{v_n(s)}&=\sum_{x\in X}\sqrt{\pi_x(s)}\ket{x}\\
            &=\sqrt{\dfrac{(1-s)(1-p_M)}{1-s(1-p_M)}}\ket{U}+\sqrt{\dfrac{p_M}{1-s(1-p_M)}}\ket{M},
\end{align} 
where $\ket{U}$ and $\ket{M}$ are defined as
\begin{align}
\ket{U}&=\frac{1}{\sqrt{1-p_M}}\sum_{x\notin M}\sqrt{\pi_x}\ket{x}\label{eqmain:stateU}\\
\ket{M}&=\frac{1}{\sqrt{p_M}}\sum_{x\in M}\sqrt{\pi_x}\ket{x}.\label{eqmain:stateM}
\end{align}  
\subsection{Hitting time and extended hitting time}\label{sec:hittingtimes}
The hitting time of a Markov chain $P$ with respect to a set of marked elements $M$ can be expressed as 
\begin{equation}
\label{eqmain:hitting_time}
HT(P,M)=\sum_{j=1}^{n-|M|}\dfrac{|\braket{v'_j}{U}|^2}{1-\lambda'_j},
\end{equation}
where $\lambda'_j$ and $\ket{v'_j}$ are the eigenvalues and eigenvectors of the matrix $D(P')$ with $\ket{U}$ defined in Eq.~\eqref{eqmain:stateU}. The hitting time is the expected number of steps needed for a classical random walk on $P$ to find one of the marked nodes in $M$, starting from a random position sampled from the stationary probability distribution $\pi$.

Furthermore, in Ref.~\cite{krovi2016quantum}, the authors define a quantity  known as the \textit{interpolated hitting time}. For an interpolated Markov chain $P(s)$, this is defined as
\begin{equation}
\label{eqmain:interpolated-hitting-time}
HT(s)=\sum_{j=1}^{n-1}\dfrac{|\braket{v_j(s)}{U}|^2}{1-\lambda_j(s)}.
\end{equation}
Taking the limit when $s\rightarrow 1$ we obtain the \textit{extended hitting time}
\begin{equation}
\label{eqmain:extended-hitting-time}
HT^+(P,M)=\lim_{s\rightarrow 1} HT(s).
\end{equation}
This quantity will be used to quantify the speed-up obtained via our quantum walk algorithms. 

Clearly for $|M|=1$, we have that $$HT^+(P,M)=HT(P,M).$$ However, in general for $|M|>1$, $$HT^+(P,M)\geq HT(P,M).$$ 

In Ref.~\cite{krovi2016quantum}, Krovi et al. proved an explicit relationship between $HT(s)$ and $HT^+(P,M)$. They showed that
\begin{equation}
\label{eqmain:interpolated-vs-extended-hitting-time}
HT(s)=\dfrac{p_M^2}{\left(1-s(1-p_M)\right)^2}HT^+(P,M).
\end{equation} 

\section{Constructing a Hamiltonian from a Markov Chain}\label{sec:SOformalism}
The work of Somma and Ortiz \cite{somma2010quantum} provides a mapping between an ergodic, reversible Markov chain $P$ and a quantum Hamiltonian. This construction has been used to develop search algorithms in the context of adiabatic quantum computation \cite{krovi2010adiabatic} and will be instrumental for the continuous-time quantum walk algorithms presented in our work.

The mapping is as follows. Let us consider a Hilbert space $\mathcal{H}\otimes\mathcal{H}$, where $\mathcal{H}=\mathrm{span}\{\ket{x}: x\in X \}$.
Also, let $p_{xy}$ denote the $(x,y)^{\mathrm{th}}$-entry of $P$ and let $E$ be the set of edges of $P$. Following the work of Szegedy \cite{szegedy2004quantum}, one can define a unitary $V$ acting on $ \mathcal{H} \otimes \mathcal{H} $ such that for all $x\in X$,
\begin{equation}
\label{eqmain:unitary-for-hamiltonian}
V\ket{x,0}=\sum_{y\in X}\sqrt{p_{xy}}\ket{x,y},
\end{equation}
where the state $\ket{0}$ represents a fixed reference state in $\mathcal{H}$. Let us also define the swap operator 
\begin{equation}
 S\ket{x,y}=
\begin{cases}
\ket{y,x}, & \text{if $(x,y)\in E$} \\
\ket{x,y}, & \text{otherwise}.
\end{cases}
\end{equation}
It can be seen that  
\begin{equation}\label{eqmain:block_discrimimant}
\braket{x,0|V^\dag S V}{y,0}=\sqrt{p_{yx}p_{xy}}=D_{xy}(P).
\end{equation}
In other words, if we define the projector $\Pi_0=I\otimes \ket{0}\bra{0}$, the discriminant matrix $D$ is encoded in a block of the operator $V^\dag S V$ given by $\Pi_0 V^\dag S V \Pi_0$.

The Hamiltonian is now defined as 
\begin{equation}
\label{eqmain:SO_hamiltonian}
H=i[V^\dag S V,\Pi_0].
\end{equation}
We shall now look at the spectrum of $H$ and investigate how it relates to the spectrum of the discriminant matrix $D(P)$. 
\subsection{Spectrum of $\mathbf{H}$}
\label{subsec:spectrum-somma-ortiz}
The spectrum of $H$ has been explicitly described in Ref.~\cite{krovi2010adiabatic} and we mention it here for completeness. Denoting the eigenstates of the discrimant matrix as $\ket{v_k}$  (eq.~\eqref{eqmain:discriminant-matrix-spectral}), the crucial observation to compute the spectrum of this Hamiltonian is that it has the following invariant subspaces
\begin{align}
&\mathcal{B}_k=\mathrm{span}\{\ket{v_k,0},V^\dag S V\ket{v_k,0}\},~~1\leq k\leq n-1\\
&\mathcal{B}_n=\mathrm{span}\{\ket{v_n,0}\}\\
&\mathcal{B}^\perp=(\oplus_{k=1}^n\mathcal{B}_k)^\perp.
\end{align}
This can be derived by first noting that $\braketbra{v_k,0}{V^\dag SV}{v_k,0}=\lambda_k$, which follows from Eq.~\eqref{eqmain:block_discrimimant}. 
Since $\lambda_n=1$ and $V^\dag S V$ is unitary, this implies that 
\begin{align}
V^\dag S V\Pi_0\ket{v_n,0}&=\ket{v_n,0},\\
\Pi_0 V^\dag S V\ket{v_n,0}&=\ket{v_n,0}.
\end{align}
Hence, we have that
\begin{equation}\label{eqmain:vn_eigenstate}
H\ket{v_n,0}=0,
\end{equation}
i.e.\ $\ket{v_n,0}$ is an eigenstate of $H$ with eigenvalue $0$ (and thus a unidimensional invariant subspace).

On the other hand, note that for $1\leq k \leq n-1$ we have 
\begin{align}
V^\dag SV\Pi_0\ket{v_k,0}&=\lambda_k\ket{v_k,0}+\sqrt{1-\lambda_k^2}\ket{v_k,0}^\perp,\\
\Pi_0 V^\dag SV\ket{v_k,0}&=\lambda_k\ket{v_k,0},\\
V^\dag SV\Pi_0\ket{v_k,0}^\perp&=0,\\
\Pi_0 V^\dag SV\ket{v_k,0}^\perp&=\sqrt{1-\lambda_k^2}\ket{v_k,0}.
\end{align}
Here, $\ket{v_k,0}^\perp$ is a quantum state that is in $\mathcal{B}_k$ such that $\Pi_0\ket{v_k,0}^\perp=0$. From this, we obtain that
\begin{align}
H\ket{v_k,0}&=i\sqrt{1-\lambda_k^2}\ket{v_k,0}^\perp\\
H\ket{v_k,0}^\perp&=-i\sqrt{1-\lambda_k^2}\ket{v_k,0},
\end{align}
which means that $H$ acts as the Pauli matrix $\sigma_y$ between $\ket{v_k,0}$ and $\ket{v_k,0}^\perp$, showing that each of the subspaces $\mathcal{B}_k$, for $1\leq k \leq n-1$, is invariant under the action of $H$. This also implies that the eigenstates and eigenvalues of $H$ in each of these subspaces are
\begin{equation}\label{eqmain:spectrumSOham}
\ket{\Psi^\pm_k}=\dfrac{\ket{v_k,0}\pm i\ket{v_k,0}^\perp}{\sqrt{2}},~~E^\pm_k=\pm \sqrt{1-\lambda_k^2}.
\end{equation}
This analysis gives us $2n-1$ out of the $n^2$ eigenvalues of $H$. It can be seen that the remaining $(n-1)^2$ eigenvalues, corresponding to the eigenvectors from the invariant subspace $\mathcal{B}^\perp$, are all $0$. However, this subspace is not relevant in the subsequent analysis of our quantum search algorithms, since we choose an initial that has no support on $B^\perp$. Thus, throughout the evolution under our search Hamiltonians the dynamics is restricted to the subspace $\mathcal{B}=\oplus_{k=1}^n\mathcal{B}_k$.

Finally, it is important to remark that this construction of $H$ ensures that the spectral gap between the $0$ eigenvalue of $H$, which encodes the stationary state of $P$, and the rest of its eigenvalues is given by
\begin{equation}
\sqrt{1-\lambda_{n-1}^2}=\Theta(\sqrt{\Delta}),    
\end{equation}
where $\Delta$ is the spectral gap of $D(P)$ (and also of $P$). This amplification of the spectral gap is crucial for our subsequent analysis of the speed-up obtained for the problem of finding marked nodes via quantum search.
\subsection{Quantum walk on the edges of a Markov chain}\label{sec:QWedges}
Although the previous analysis shows that the spectrum of $H$ is related to that of $D(P)$ (and in turn $P$), the locality of $H$ is not clear from its definition and has not been analysed explicitly in the previous works using the Somma-Ortiz construction. 

Here, we demonstrate that the dynamics under a rotated version of $H$ can be seen as a quantum walk on the edges of $P$.  Let us the define this rotated Hamiltonian as
\begin{equation}
\label{eqmain:rotated-hamiltonian}
\overline{H}=V H V^\dag=i[S,V\Pi_0V^\dag].
\end{equation}
Each entry of $\overline{H}$ is given by
\begin{equation}
\label{eq:ham_entry}
\braket{x',y'|\overline{H}}{x,y}=i\left( \delta_{x,y'}\sqrt{p_{y'x'}p_{xy}}-\delta_{x',y}\sqrt{p_{x'y'}p_{yx}}\right),
\end{equation}
where $\delta_{xy}$ is the Kronecker-delta function. The situations in which this matrix element is non-zero can be reduced to the following cases
\begin{itemize}
\item[(i)]~ If $\{(y,x), (y,z)\}\in E$ with $x\neq z$, we have 
\begin{equation}\label{eqmain:connectivity}
 \braket{y,z|\overline{H}}{x,y}=-i\sqrt{p_{yx}p_{yz}};   
\end{equation}
\item[(ii)] If $\{(x,y), (y,x)\}\in E$, we have
\begin{equation}
\braket{y,x|\overline{H}}{x,y}=i\left(p_{xy}-p_{yx}\right).
\end{equation}
\end{itemize}
The other two cases can be obtained by complex conjugation of the previous equations.
Hence, if the walker is localized in a directed edge from node $x$ to node $y$, i.e.\ $\ket{x,y}$, then it can move to a superposition of outgoing edges from node $y$ of the form $\ket{y,.}$. A similar connectivity can be obtained for the Szegedy walk operator $U=S(2 V \Pi_0 V^{\dagger} - I)$, which defines a discrete-time quantum walk on a Markov chain.  

From the definition of $\overline{H}$, it is clear that its eigenvalues are the same as those of $H$ and its eigenstates can be obtained by rotating the eigenstates of $H$ with unitary $V$. In particular, the eigenstate of eigenvalue $0$ of $\overline{H}$ is
\begin{equation}
\ket{\overline{v}_n}=V\ket{v_n,0}=\sum_{x\in X}\sqrt{\pi_x}V\ket{x,0}=\sum_{(x,y)\in E}\sqrt{\pi_x p_{xy}}\ket{x,y}.
\end{equation}
For example, if $P$ represents the transition matrix of a random walk on a simple graph with a set of edges $E$,  we have $p_{xy}=A_{xy}/d_x$, where $A_{xy}$ are the entries of the adjacency matrix and $d_x$ is the degree of the node $x$. Also, it is well known that the $x^{\mathrm{th}}$-entry of the stationary state $\pi$ of such graphs is $\pi_x=d(x)/|E|$. Then it is easy to verify that for all such graphs, 
\begin{equation}
\ket{\overline{v}_n}=\dfrac{1}{\sqrt{|E|}}\sum_{(x,y)\in E}\ket{x,y},
\end{equation}
i.e.\ it is the equal superposition of all the edges of the underlying graph.

In conclusion, given an ergodic reversible Markov chain $P$, the evolution under  $\overline{H}$ defines a continuous-time quantum walk on the edges of $P$. However, in the subsequent analysis of our algorithms, we shall be working with $H$ as it simplifies some of the calculations. Our results can be directly applied to search algorithms involving $\overline{H}$ by suitable rotations of the initial and final state with unitary $V$.
\section{Childs and Goldstone algorithm for any ergodic, reversible Markov chain} 
\label{sec-main:modified-cg-algo} 
In this section we present a modified version of the spatial search algorithm proposed by Childs and Goldstone ($\C\G$ algorithm) which can be used to search for a marked node in any ergodic, reversible Markov chain. Before we introduce this algorithm, which we refer to as $\C\G'$, we start by briefly describing the approach used in previous works for doing quantum search by continuous-time quantum walk.
\subsection{The Childs and Goldstone algorithm for spatial search}
We begin by stating the general framework of the $\C\G$ algorithm. Consider a graph $G$ with a set of $n$ vertices labelled $\{1,2,...n\}$ and a Hamiltonian $H_G$ that encodes the connectivity of the underlying graph, defined on a Hilbert space $\mathcal{H}$ of dimension $n$. In previous applications of this algorithm, the graph is usually an undirected graph, and $H_G$ is taken to be proportional to the adjacency matrix of the graph or the graph's Laplacian matrix. In this framework, the search Hamiltonian is given by
\begin{equation}
\label{eqmain-search-ham-def}
H_{\mathrm{search}}=H_{\mathrm{oracle}}+ r H_G,
\end{equation}
where $r$ is a (non-zero) tunable parameter and $H_\mathrm{oracle}$ is the oracular Hamiltonian that singles out the marked node, which we shall denote as $\ket{w}$ (for the moment let us consider that a single node is marked). We also require that $H_{\mathrm{oracle}}$ is local so that it perturbs the node $\ket{w}$ in a way that affects only vertices (or edges) in its vicinity. For example, the most widely used version of the $\C \G$ algorithm considers that $H_{\mathrm{oracle}}=\ket{w}\bra{w}$, which adds a local energy at node $\ket{w}$, leaving the remaining vertices unaffected. In fact, simulating this oracular Hamiltonian for a time $t$, corresponds to $\Oo(t)$-queries to the oracle of the Grover's search algorithm \cite{roland2003quantum}. The algorithm then involves choosing an appropriate value of $r$ such that evolving $H_{\mathrm{search}}$ for some time $T$, starting from a state containing no information about $\ket{w}$, leads to some state $\ket{f}$ that has a good overlap with $\ket{w}$.

Alternatively, one can also use an oracle that affects the edges in the vicinity of $\ket{w}$ of the form 
\begin{equation}\label{eqmain:oracle_edges}
    H_{\mathrm{oracle}}=-H_G\ket{w}\bra{w}-\ket{w}\bra{w}H_G. 
    \end{equation}
The search Hamiltonian, defined as
\begin{equation}\label{eqmain:searchCG}
    H_{\mathrm{search}}=H_G-H_G\ket{w}\bra{w}-\ket{w}\bra{w}H_G, 
\end{equation}
is of the form of Eq.~\eqref{eqmain-search-ham-def} with $H_{\mathrm{oracle}}$ from Eq.~\eqref{eqmain:oracle_edges} and $r=1$. A similar framework was considered, for example, in Refs.~\cite{foulger2014quantum, childs2014spatial-crystal} to analyse quantum search on graphene and crystal lattices. 

In this scenario, $\bra{w}H_G\ket{w}=0$ and so this search Hamiltonian decouples the marked node $\ket{w}$ from the rest of the Hilbert space, since $H_{\mathrm{search}}\ket{w}=0$. The dynamics is such that, after an certain amount of time, the wavefunction has a large overlap with the state 
\begin{equation}
    \ket{\tilde{w}}=\frac{H_G\ket{w}}{||H_G\ket{w}||}, 
\end{equation}
which is a superposition of the states that are directly coupled to $\ket{w}$ via $H_G$. The state $\ket{w}$ can be prepared from $\ket{\tilde{w}}$ by evolving this state under $e^{i H_{\mathrm{oracle}} t'}$ in time $t'=O(1/||H_G\ket{w}||)$,  since $H_{\mathrm{oracle}}$ generates a rotation in the subspace $\text{span}\{\ket{w},\ket{\tilde{w}}\}$.
\subsection{Modifying the Childs and Goldstone algorithm for searching Markov chains}
Inspired by the latter approach to quantum search, we propose a modification of Childs and Goldstone algorithm to encompass quantum search on any ergodic reversible Markov chain $P$, which we refer to as $\C\G'$ algorithm. The main idea is to consider a larger Hilbert space $\mathcal{H}\otimes \mathcal{H}$ of dimension $n^2$ and use the formalism of Somma-Ortiz, reviewed in Sec.~\ref{sec:SOformalism}, to construct from $P$ the Hamiltonian driving the quantum walk. 

In this enlarged Hilbert space, we denote the marked node as $\ket{w, 0}$ and define the oracle Hamiltonian and the search Hamiltonian as
\begin{align}
    H_{\mathrm{oracle}}&=-H\ket{w,0}\bra{w,0}-\ket{w,0}\bra{w,0}H,\label{eqmain:oracleCGprime}\\
    H_{\mathrm{search}}&=H-H\ket{w,0}\bra{w,0}-\ket{w,0}\bra{w,0}H, \label{eqmain:HsearchCGprime}
\end{align}
where $H=i [V^{\dagger}SV,\Pi_0]$ as defined in Eq.~\eqref{eqmain:SO_hamiltonian}. Note that this search Hamiltonian has a similar form to that of Eq.~\eqref{eqmain:searchCG}. Importantly, it can be verified that $\bra{w,0}H\ket{w,0}=0$ and hence this search Hamiltonian also decouples the marked state from the rest of the Hilbert space as $H_{\mathrm{search}}\ket{w,0}=0$. 

Before we present the $\C\G'$ algorithm, we introduce a parameter that is crucial to understand the running time of this approach to quantum search. If $  \lambda_i$ and $\ket{v_i}$ are the eigenstates of the discriminant matrix $D(P)$ (see Eq.~\eqref{eqmain:discriminant-matrix-spectral}),  we define the parameter 
\begin{equation}
\label{eq-main:mu}
\mu=\sqrt{\sum_{i=1}^{n-1}\dfrac{2|a_i|^2}{1-\lambda_i^2}},
\end{equation}
where $a_i=\braket{w}{v_i}$. Furthermore, recall that the state $\ket{v_n,0}$ is an eigenstate of $H$ with eigenvalue $0$ (see Eq.~\eqref{eqmain:vn_eigenstate}). We choose this state as the initial state of the $\C\G'$ algorithm, similar to DTQW-based spatial search algorithms \cite{szegedy2004quantum,krovi2016quantum}, and denote its initial overlap with $\ket{w,0}$ as 
\begin{equation}
\sqrt{\epsilon}=|\braket{w,0}{v_n,0}|=|a_n|.    
\end{equation}

The steps of the $\C\G'$ algorithm are described in Algorithm~1 below. This algorithm prepares a quantum state $\ket{f}$ with an overlap with the solution state given by $\nu=|\braket{w,0}{f}|$.  
\RestyleAlgo{boxruled}
\begin{algorithm}[h!]
  \caption{$\C\G'$ algorithm}\label{algo-cg-modified}
  Given an ergodic, reversible Markov chain $P$:
  \begin{itemize}
  \item[1.] Prepare the $0$-eigenstate of $H=i [V^{\dagger}SV,\Pi_0]$, i.e.\ $\ket{v_n,0}$.\\
  \item[2.] Evolve this state under $H_{\mathrm{search}}$ from Eq.~\eqref{eqmain:HsearchCGprime} for time $$T_1=\frac{\pi}{2}\frac{\mu}{\sqrt{\epsilon}}$$ to obtain the state $\ket{\tilde{f}}$.
  \item[3.] Evolve $\ket{\tilde{f}}$ under the action of $H_{\mathrm{oracle}}$ from Eq.~\eqref{eqmain:oracleCGprime} for time $T_2=\frac{\pi}{2||H\ket{w}||}$.   \end{itemize}
  \end{algorithm}
In order to provide general bounds for this overlap, we need to impose a condition on the spectrum of the discriminant matrix $D(P)$, which determines the regime of validity of our perturbative analysis. This condition can be written in terms of the parameters $\mu$, the initial overlap $\sqrt{\epsilon}$ and the spectral gap between the two highest eigenvalues of $D(P)$ as     
 \begin{equation}\label{eqmain:spec_con}
    \sqrt{\epsilon}\leq c \sqrt{\Delta}\mu,  
 \end{equation}
 where $c$ is a small positive constant. Note that usually the initial overlap $\sqrt{\epsilon}$ is quite small: for example, for state-transitive Markov chains we have $\epsilon=1/\sqrt{n}$. 
 
 Also, it can be seen from Eq.~\eqref{eq-main:mu} that $\mu\geq \sqrt{2-2\epsilon}$ (we recall that, for simplicity, we work with a shifted Markov chain $P\rightarrow (I+P)/2$ so that $\lambda_i\geq 0$). This ensures that our analysis is valid, for example, when $\Delta\gg \epsilon$. 
 
 Our main result regarding the performance of the $\C\G'$ algorithm is the following. 
 \\~\\
 \begin{restatable}{theorem}{searchcgprime}
\label{thm-main:search-somma-ortiz}
Let $P$ be an ergodic reversible Markov chain, whose corresponding discriminant matrix $D(P)$ fulfills the spectral condition from Eq.~\eqref{eqmain:spec_con} for a given marked state $\ket{w}$. Then Algorithm \ref{algo-cg-modified} outputs a quantum state $\ket{f}$ that has an overlap with $\ket{w,0}$ of 
\begin{equation}\label{eqmain:amplitudeCGprime}
\nu=|\braket{w,0}{f}|=\Theta\left(\frac{1}{\mu\nrm{H\ket{w,0}}}\right)
\end{equation}
in time
\begin{equation}\label{eqmain:timeCGprime}
T=T_1+T_2=\Theta\left(\dfrac{\mu}{\sqrt{\epsilon}}\right).    
\end{equation}
\end{restatable}

\textbf{Proof:} See Sec.~\ref{sec:proof-search-somma-ortiz} of the Appendix.
\\~\\
We will see in Sec.~\ref{sec:performanceCGprime} that from this result we can derive that the $\C\G'$ algorithm has an improved performance over the original approach by Childs and Goldstone for several important examples. 

The main steps to demonstrate this result are as follows. First, we show via Lemma~\ref{lem:optimal-search-symmetric-spectrum} in the Appendix that step 2 of the algorithm prepares a state $\ket{\widetilde{f}}$ with an overlap $\nu$ (Eq.~\eqref{eqmain:amplitudeCGprime}) with the state 
\begin{equation}
\label{eqmain:final-evolution-state-modified-cg}
\ket{\widetilde{w}}=\dfrac{H\ket{w,0}}{\nrm{H\ket{w,0}}}, 
\end{equation}
which is a superposition of the states directly coupled to the solution $\ket{w,0}$ via Hamiltonian $H$. Precisely, this state has the form
\begin{equation}
\ket{\widetilde{f}}=\nu \ket{\widetilde{w}}+\sqrt{\epsilon}\ket{w,0}+\ket{\widetilde{w}}^\perp,
\end{equation}
where $\ket{\widetilde{w}}^\perp$ is an (unnormalized) quantum state which is orthogonal to both $\ket{w,0}$ and $\ket{\tilde{w}}$.

Finally, in step 3 of the algorithm, the evolution under $H_{\mathrm{oracle}}$ for time $T_2=\pi/2\nrm{H\ket{w,0}}^{-1}$ generates a rotation between $\ket{\tilde{w}}$ and $\ket{w,0}$ which leads to a final state $\ket{f}$ with an overlap of $\nu$ with the solution state. Importantly, we can show the upper bound $T_2= O(\mu)$ which implies that $T_2$ is significantly lower than $T_1$.  This can be seen from the fact that
\begin{equation}
\nrm{H\ket{w,0}}=\sqrt{\braket{w,0}{H^2|w,0}}=\sqrt{\sum_{i= 1}^{n-1}2|a_i|^2(1-\lambda^2_i)},    
\end{equation}
which follows from the spectral properties of $H$ derived in Sec.~\ref{subsec:spectrum-somma-ortiz}. Using the Cauchy-Schwarz inequality, we obtain that
\begin{align}\label{eqmain:bound1}
\mu \nrm{H\ket{w,0}}&=2\sqrt{\left(\sum_{i=1}^{n-1}\dfrac{|a_i|^2}{1-\lambda^2_i}\right)\left(\sum_{i=1}^{n-1}|a_i|^2(1-\lambda^2_i)\right)} \\
                    &\geq 2 \sum_{i=1}^{n-1} |a_i|^2=2(1-\epsilon)\label{eqmain:bound2}.
\end{align}
Hence, $T_2=O(\nrm{H\ket{w,0}}^{-1})=O(\mu)$, implying that the total evolution time is dominated by $T_1$, i.e., $T=\Theta(T_1)=\Theta(\mu/\epsilon)$.

After obtaining $\ket{f}$, a measurement in the basis spanned by the state-space of $P$, post-selected on having $\ket{0}$ in the second register, gives us the solution node with probability $\nu^2$. Alternatively, the solution node can also be obtained from $\ket{f}$ by using $\Theta(1/\nu)$-rounds of amplitude amplification.

It is worth noting that in order to compute the complexity of Algorithm \ref{algo-cg-modified}, we have ignored (i) the cost of preparing the initial state $\ket{v_n,0}$, known as the setup cost and (ii) the cost of measuring in the basis spanned by the state-space of the Markov chain. Also, we have assumed that the cost of simulating the Hamiltonian $H$ for unit time is constant. For details on how these costs impact the overall running time of a CTQW-based algorithm, we refer the reader to Ref.~\cite{shanleonew2019}.

\subsection{Performance of the $\C\G'$ algorithm}\label{sec:performanceCGprime}
We can now analyse the running time of the $\C\G'$ algorithm for some important examples and demonstrate some advantages with respect to the original Childs and Golstone algorithm. We recall that the latter uses a search Hamiltonian of the form of Eq.~\eqref{eqmain-search-ham-def} with $H_{\mathrm{oracle}}=-\ket{w}\bra{w}$. To do this comparison, we will focus on random walks on undirected graphs, which can be seen as a Markov process with transition matrix $p_{xy}=A_{xy}/d_x$, where $A_{xy}=A_{yx}$ is the adjacency matrix of an undirected graph and $d_x$ is the degree of the vertex $x$. As before, we restrict ourselves to the case where $P$ is ergodic and reversible. In this scenario, it is natural to choose the driving Hamiltonian $H_G$ from Eq.~\eqref{eqmain-search-ham-def} to be the normalized adjacency matrix of the graph with the $(x,y)^{\mathrm{th}}$ entry being $A_{xy}/\sqrt{d_x d_y}$. This is exactly the discriminant matrix $D(P)$ (see Eq.~\eqref{eqmain:discriminant-matrix-definition}). 
An important advantage of $\C\G'$ algorithm over the $\C\G$ algorithm is that while the latter needs a careful tuning of the hopping strength $r$ from Eq.~\eqref{eqmain-search-ham-def} which is dependent on the spectrum of the underlying graph, the $\C\G'$ algorithm simply sets $r=1$. In what follows, we provide examples of instances where the $\C\G'$ algorithm performs better than the $\C\G$ algorithm and also elucidate on some of the drawbacks of this approach.  
\subsubsection{$d$-dimensional lattices}
For $d$-dimensional lattices with $d\geq 2$, the spectral gap of $D(P)$ scales as $n^{-2/d}$. A classical random walk on $P$ has hitting time in $\Oo(n)$ if $d\geq 3$ while for $d=2$, the hitting time is in $\Oo(n\log n)$. The original CTQW-based spatial search algorithm by Childs and Goldstone, the running time is $\Oo(\sqrt{n})$ for lattices of $d>4$, while for $d=4$, the running time is in $\Oo(n\log n)$, while the there is no substantial speedup for $d<4$.

On the other hand for the $\C\G'$ algorithm, for dimension $d\geq 3$, the solution node is found with constant probability after an the evolution time $T=\Theta(\sqrt{n})$. This can be seen by noting that noting that 
\begin{equation}\label{eqmain:boundmu}
    \sqrt{S_1}\leq \mu \leq \sqrt{2 S_1}, 
\end{equation}
with 
\begin{equation}\label{eqmain:S1}
    S_1=\sum_{i=1}^{n-1}\frac{|a_i|^2}{1-\lambda_i}.
\end{equation}
The latter parameter was computed for lattices in Ref.~\cite{childs2004spatial} and is given by $S_1=\Theta(\log(n))$ for $d=2$ and $S_1=\Theta(1)$ for $d\geq 3$.
From Theorem~\ref{thm-main:search-somma-ortiz}, this implies that for $2d$-lattices the $\C\G'$ algorithm can reach an amplitude at the marked node of $\Theta(1/\sqrt{\log n})$ in time $T=\Theta(\sqrt{n\log n})$. Hence,  the solution node can be obtained in $\Theta(\sqrt{n}\log n)$ by using $\Theta(\sqrt{\log n})$-rounds of amplitude amplification. This running time is better than the original $\C\G$ algorithm and matches the performance of the CTQW algorithm where the walker has an additional spin degree of freedom \cite{childs2004spatial-dirac}. However, this is still slower than the square root of the classical hitting time, albeit by a factor of $\Theta(\sqrt{\log n})$. 
\subsubsection{State-transitive graphs}
Here we demonstrate that for any state-transitive graph with a hitting time of $HT=\Theta(n)$, the $\C\G'$ algorithm provides a quadratic speed-up over the corresponding classical random walk  (provided they satisfy the spectral condition from Eq.~\eqref{eqmain:spec_con}). Note that for state-transitive graphs this condition can be simplified to $\Delta\gg 1/n$ (note that $\mu\geq 1$ and $\epsilon=1/n$ for state-transitive graphs). 

To demonstrate this, we note that if $P$ is state transitive, its hitting time with respect to the marked element $w$ is given by 
\begin{equation}\label{eqmain:HTandS1}
HT(P,w)=n S_1,    
\end{equation}
with $S_1$ defined in \eqref{eqmain:S1} \cite{szegedy2004quantum}. Hence, for state-transitive where Markov chains with $HT(P,w)=\Theta(n)$ we have from Eqs.~\eqref{eqmain:HTandS1} and \eqref{eqmain:boundmu} that $\mu=\Theta(1)$. Hence, from Theorem~\ref{thm-main:search-somma-ortiz}, we see that Algorithm~\ref{algo-cg-modified} prepares a state with constant amplitude after a time $O(\sqrt{n})$, providing a quadratic speed-up with respect to the classical hitting time. 

Note that for the original $\C\G$ algorithm such a general quadratic speed-up is not possible. It was demonstrated in Ref.~\cite{shanleonew2019} that this algorithm fails to achieve a quadratic speed-up for an unweighted Rook's graph, a graph whose connectivity is related to the possible movements of a rook on a rectangular chessboard. This is a state-transitive graph whose hitting time is $\Theta(n)$.  For certain proportions of the chessboard, the maximum amplitude at the marked node using the $\C\G$ algorithm can be as low as $n^{-1/8}$ and is reached in an evolution time $T=\Theta(n^{5/8})\gg n^{1/2}$. Hence, for this family of graphs, the performance of the $\C\G'$ algorithm can be significantly better when compared to the original Childs and Goldstone approach.  
\subsubsection{Worst-case performance and open questions}
Given the previous examples, one might wonder whether the $\C\G'$ algorithm can always provide a quadratic speed-up (up to log factors) with respect to the classical hitting time of a random walk on $P$. Unfortunately, this does not seem to be the case as we can show by analysing the worst case performance of the $\C\G'$ algorithm predicted by Theorem~\ref{thm-main:search-somma-ortiz}.

Assuming the condition $\sqrt{\epsilon}\leq c\mu\sqrt{\Delta}$ is satisfied, we have that the maximum running time for a given Markov chain $P$ is given by 
\begin{equation}
\label{eqmain:upper-bound-evolution-time}
T=\mathcal{O}\left(\dfrac{1}{\sqrt{\Delta\epsilon}}\right),
\end{equation}
where we use the following bound on $\mu$
\begin{equation}
\mu=\sqrt{\sum_{i=1}^{n-1}\dfrac{2|a_i|^2}{1-\lambda_i^2}}\leq \sqrt{\dfrac{2(1-\epsilon)}{\Delta}}.    
\end{equation}
Furthermore, from the upper bound $||H\ket{w}||\leq 1$,  the final overlap with the marked node can be as low as \begin{equation}\label{eqmain:amplitudelowerbound}
    \nu=\Theta(\sqrt{\Delta}).
\end{equation}
In Appendix~\ref{sec:worstcaseCGprime}, we demonstrate an example of a weighted Rook's graph, for which the upper bound on the time from Eq.~\eqref{eqmain:upper-bound-evolution-time} and the lower bound on the amplitude from Eq.~\eqref{eqmain:amplitudelowerbound} are simultaneously attained. Note that in general, as the gap $\Delta$ can be a decreasing function of $n$, the $\C\G'$ algorithm fails to achieving a generic quadratic speed-up with respect to the hitting time of a classical random walk on any ergodic, reversible Markov chain $P$, which can be upper bounded by 
$HT(P,\{w\})=\mathcal{O}\left(\frac{1}{\Delta\epsilon}\right)$ \cite{szegedy2004quantum}. Interestingly, for this example, the performance of the $\C\G'$ algorithm predicted via Theorem~\ref{thm-main:search-somma-ortiz} seems to be considerably worse than that of the original Childs and Goldstone approach.

It is thus natural to ask whether simple modifications of the $\C\G'$ by for example, introducing a tunable parameter $r$ controlling the weight of the terms $H$ and $H_{oracle}$ for the search Hamiltonian, similarly to \eqref{eqmain:searchCG} (in the current approach we take $r=1$ in the second step and $r=0$ in the third). We believe techniques similar to those employed in Refs.~\cite{shanleonew2019, childs2004spatial} could be used to show that the current approach is the best possible one, but leave an explicit demonstration of this fact as an open question.

Overall, although the $\C\G'$ algoritm has some attractive features, namely, it can be applied to any ergodic, reversible Markov chain and does not require any parameter to be tuned,  it fails to achieve a general quadratic speed-up with over classical random walks. Furthermore, it seems difficult to obtain general results regarding the performance of this approach in the presence of multiple solutions. Next we show that these problems can be surmounted by exploiting the framework of interpolated Markov chains.

\section{Quantum spatial search on interpolated Markov chains} 
\label{sec-main:search-algo-phase-randomization}
In this section, we provide a spatial search algorithm that finds a marked node on any ergodic, reversible Markov chain in square root of the classical hitting time. In the scenario where multiple nodes are marked, our algorithm solves this problem in square root of the extended hitting time. 

The main idea is to construct a search Hamiltonian from an interpolated Markov chain, introduced in Sec.~\ref{sec:interpolatedMC}, via the Somma-Ortiz formalism reviewed in Sec.~\ref{sec:SOformalism}. Precisely, for a given ergodic reversible Markov chain $P$ with a marked set of nodes $M$, we consider the interpolated Markov chain  
\begin{equation}
P(s)=(1-s)P+sP',     
\end{equation}
where $P'$ is obtained from $P$ by removing all the outgoing edges from $M$ and replacing them by self loops (see Eqs.~\eqref{eqmain:PandPprime} and \eqref{eqmain:Pofs}). We recall that for any 
$0\leq s < 1$, $P(s)$ is also ergodic and reversible.

From such an interpolated Markov chain, we construct a Hamiltonian as in Sec.~\ref{sec:SOformalism} (which now depends explicitly on $s$) given by   
\begin{equation}
H(s)=i[V(s)SV(s)^\dag,\Pi_0],
\end{equation}
where the unitary \footnote{$V(s)$ can be constructed by using $V(0)$ and single qubit rotations to an ancilla qubit. See Ref.~\cite{krovi2016quantum}}
\begin{equation}
V(s)\ket{x,0}=\sum_{x,y}\sqrt{p_{xy}(s)}\ket{x,y}.
\end{equation}
Note that this Hamiltonian cannot in general be written as a sum of a term encoding $P$ and an oracular term depending on the marked set $M$, as in the Childs and Goldstone formalism and its modified version discussed in Sec.~\ref{sec-main:modified-cg-algo}.  

The Hamiltonian $H(s)$ generates a quantum walk on the edges of $P(s)$ and its spectrum can be computed from the spectral properties of the discriminant matrix $D(P(s))$, as elucidated in Sec.~\ref{subsec:spectrum-somma-ortiz} and in Sec.~\ref{sec:QWedges}. Namely, we recall that the eigenstate of eigenvalue $1$ of $D(P(s))$ is given by 
\begin{equation}
\label{eqmain:re-express-v-n}
\ket{v_n(s)}=\cos\theta(s)\ket{U}+\sin\theta(s)\ket{M},
\end{equation}
where 
\begin{align}
\cos\theta(s)&=\sqrt{\dfrac{(1-s)(1-p_M)}{1-s(1-p_M)}},\label{eqmain:costheta}\\
\sin\theta(s)&=\sqrt{\dfrac{p_M}{1-s(1-p_M)}}, \label{eqmain:sintheta}    
\end{align}
with $p_M=\sum_{x\in M}\pi_x$ and $\ket{U}$ and $\ket{M}$ defined in Eqs.~\eqref{eqmain:stateU} and \eqref{eqmain:stateM}, respectively. At $s=0$ we recover the state $\ket{v_n(0)}$, which can be written as 
\begin{equation}
\label{eqmain:start_state}
\ket{v_n(0)}=\sqrt{1-p_M}\ket{U}+\sqrt{p_M}\ket{M}, 
\end{equation}
whereas at $s=1$ the state only has support over the marked subspace. 

Since $\ket{v_n(s),0}$ is an eigenstate with energy $0$ of $H(s)$, we can adiabatically evolve the state $\ket{v_n(0),0}$ under $H(s)$, by interpolating the parameter $s$ from $0$ to $1$, in order to prepare the state $\ket{M}$. It was shown in Ref.~\cite{krovi2010adiabatic} that the time for this adiabatic quantum algorithm to succeed with constant probability can be bounded in terms of the \emph{extended hitting time} defined in Eq.~\eqref{eqmain:extended-hitting-time}. In what follows, we show that the search problem can also be solved with the same running time by evolving a \emph{time-independent} Hamiltonian $H(s^*)$.
\subsection{Spatial search via randomized time evolution}\label{subsec:algorandomt}
Let us consider the value $s^*=1-p_M/(1- p_M)$. It is easy to see that 
\begin{equation}\label{eqmain:vn_sstar}
\ket{v_n(s^*)}=\dfrac{\ket{U}+\ket{M}}{\sqrt{2}},
\end{equation}
as $\cos\theta(s^*)=\sin\theta(s^*)=1/\sqrt{2}$.

This state not only has a large overlap with the marked subspace, but also with the state $\ket{v_n(0)}$ since
\begin{align}
\alpha_n&=\braket{v_n(s^*)}{v_n(0)}\\ &=\sqrt{1-p_M}\cos\theta(s^*)+\sqrt{p_M}\sin\theta(s^*)\\
&=\dfrac{\sqrt{1-p_M}+\sqrt{p_M}}{\sqrt{2}}, \label{eqmain:overlap_highest_estate_start_state}
\end{align}
where we assume $p_M$ to be small \footnote{If $p_M$ is a constant, simply preparing $\ket{v_n(0)}$ and measuring in the state space basis, yields the marked node with a constant probability and as such the interesting, non trivial case occurs when $p_M$ is small.}. 

Give this information, let us now consider the following (hypothetical) procedure:
\begin{itemize}
    \item[1)] Prepare the state $\ket{v_n(0),0}$. 
    \item[2)] Perform a projective measurement with measurement operators
    \begin{align*}
    \M_1&=\ket{v_n(s^*),0}\bra{v_n(s^*),0}\\
    \M_2&= I-\ket{v_n(s^*),0}\bra{v_n(s^*),0},
    \end{align*}
    discarding the measurement outcome.
    \item[3)] Measure in the basis spanned by the state space.
\end{itemize} 
The repetition of this process for a constant number of times would be able to find a marked state with high probability. This can be seen by noting that after step 2, the state is diagonal in the eigenbasis of $H(s^*)$ and has the form 
\begin{align}
\begin{split}
\rho=&|\alpha_n|^2\ket{v_n(s^*),0}\bra{v_n(s^*),0}+ \\
				  &\sum_{k,l=1}^{n-1}\sum_{\sigma,\sigma' =\pm}\alpha_k\alpha_l^*  \ket{\Psi^{\sigma}_k(s^*)}\bra{\Psi^{\sigma'}_l(s^*)}, 
\end{split}
\end{align}  
where $\ket{\Psi^{\sigma}_k(s^*)}$ are eigenstates of $H(s^*)$ given in Eq.~\eqref{eqmain:spectrumSOham} and $\alpha_k= \braket{\Psi^{\sigma}_k(s^*)}{v_n(0),0}$. Using Eqs.~\eqref{eqmain:vn_sstar} and \eqref{eqmain:overlap_highest_estate_start_state}, we conclude that a measurement of the first register of this state in the state-space basis would find a marked state with probability 
\begin{equation}
    \mathrm{Tr}\left[(\Pi_M\otimes I)\rho\right]\geq \frac{|\alpha_n|^2}{2}\geq \frac{1}{4}, 
\end{equation}
where $\Pi_M=\sum_{x\in M}\ket{x}\bra{x}$. 

However, step 2 cannot be implemented directy since, for example, we cannot assume we have access to measurements on the eigenbasis of $H(s^*)$. Nevertheless, we can \emph{dephase} the state $\ket{v_n(0),0}$ in the eigenbasis of $H(s^*)$ via a technique known as \textit{quantum phase randomization} \cite{boixo2009eigenpath} (for details see Sec.~\ref{sec:quantum-phase-randomization} of the Appendix). 

The key idea is to consider the evolution of the state $\ket{v_n(0),0}$ with Hamiltonian $H(s^*)$ for a time $t\in [0,T]$ chosen \emph{uniformly at random}. The expected quantum state after this random time evolution can be written as 
\begin{align}
\label{eq:final_density_matrix}
\begin{split}
&\overline{\rho}(T)=\dfrac{1}{T}\int_0^T dt~  e^{-i H(s^*)t}\ket{v_n(0),0}\bra{v_n(0),0} e^{i H(s^*)t} \\
                  &=|\alpha_n|^2\ket{v_n(s^*),0}\bra{v_n(s^*),0}+\rho'(T)\\&+ \sum_{k=1}^{n-1}\sum_{\sigma=\pm}\left(\alpha_j^* \alpha_n\dfrac{e^{i  E_j^\sigma T}-1}{i E_j^\sigma T} \ket{v_n(s),0}\bra{\Psi^{\sigma}_j(s)} +\text{h.c.}\right)
\end{split}
\end{align}
where $\rho'(T)$ is a state with support only on states orthogonal to $\ket{v_n(s^*),0}$.
Note that by increasing $T$, we can decrease the strength of the off-diagonal elements of the time-averaged density matrix of the form $\ket{v_n(s),0}\bra{\Psi^{\sigma}_j(s)}$ and its Hermitian conjugate. In fact, we demonstrate that if $T$ is sufficiently larger than the square root of the Extended Hitting Time, these terms play a negligible role and we can lower bound the probability of finding a marked element by measuring $\overline{\rho}(T)$ in the state-space basis with a value close to $1/4$. 

The steps of the spatial search algorithm we propose are detailed in Algorithm~\ref{algo1}. Our main result regarding the performance of this algorithm is the following.
\RestyleAlgo{boxruled}
\begin{algorithm}[ht]
  \caption{Quantum spatial search by quantum phase randomization}\label{algo1}
Consider the Hamiltonian $H(s)=i[V(s)^\dag S V(s), \Pi_0]$.  
  \begin{itemize}
  \item[1.]~Prepare the state $\ket{v_n(0),0}$.\\
  \item[2.]~For $s^*=1-p_M/(1-p_M)$, $\eps\in(0,1/4)$ and $T=\Theta(\frac{1}{\eps}\sqrt{HT^+(P,M)/2})$, evolve according to $H(s^*)$ for a time chosen uniformly at random between $[0,T]$.\\
  \item[3.]~Measure in the basis spanned by the state space, in the first register.
  \end{itemize}
\end{algorithm}
\begin{theorem}
\label{thm-main:search-phase-randomization}
For any ergodic, reversible Markov chain $P$ with a set of marked nodes $M$, Algorithm \ref{algo1} finds a marked node in $M$ with an expected probability 
$$\mathrm{Tr}\left[(\Pi_M\otimes I)\bar{\rho}(T)\right]\geq 1/4-\eps,$$
provided 
$$T\geq \dfrac{1}{\eps}\sqrt{\dfrac{HT^+(P,M)}{2}},$$
where $\Pi_M=\sum_{x\in M}\ket{x}\bra{x}$ and $HT^+(P,M)$ is the extended hitting time of $P$ with respect to $M$.  
\end{theorem}
\textbf{Proof:} See Sec.~\ref{sec:proof-search-phase-randomization} of the Appendix.
\\~\\
As shown in Sec.~\ref{sec:hittingtimes}, when a single node is marked, i.e.\ $|M|=1$, $HT(P,M)=HT^+(P,M)$ and so a full quadratic speedup over the hitting time of classical random walks is obtained in this scenario. However for $|M|>1,~HT^+(P,M)\geq HT(P,M)$ and hence the problem of whether a full quadratic speedup is possible in the case of multiple marked vertices is still open in this framework.

This discrepancy in the running time of the spatial search algorithm also existed in the discrete-time quantum walk framework until recently. After the preparation of the first version of this manuscript, Ambainis et al. \cite{ambainis2019quadratic}, using the framework of interpolated quantum walks, provided a quantum algorithm with a full quadratic speedup for this problem (up to a logarithmic overhead) even in the scenario where multiple nodes are marked. It would be interesting to explore whether the same is possible for CTQW-based spatial search.   

Interestingly, we can relate the CTQW-based framework we presented in this article to its discrete-time counterpart. In particular, the Hamiltonian $H(s)$ can be simulated efficiently using only query access to the DTQW unitary, $W(s)$ of Refs.~\cite{krovi2016quantum, ambainis2019quadratic}, i.e.\ $e^{-itH(s)}$ can be simulated to $\eps$-precision by $\Oo(t+\log 1/\eps)$-queries to $W(s)$ (Please refer to Appendix \ref{sec:quantum-simulation} for details). This connection can be explored to design novel CTQW-based algorithms which have, until now, been developed only in the DTQW-based framework. For example, we believe that this connection, in conjunction with the recent results of Ambainis et al.~\cite{ambainis2019quadratic} can lead to a CTQW-based spatial search algorithm with a full quadratic speedup, even when multiple nodes are marked. 

\section{Discussion} 
In this article, we have resolved several long-standing differences between spatial search by discrete-time quantum walk (DTQW) and continuous-time quantum walk (CTQW). DTQW-based spatial search algorithms can find a marked node on any ergodic, reversible Markov chain in the square root of the classical hitting time. On the other hand, the only previously known CTQW-based algorithm (by Childs and Goldstone, denoted as $\C\G$ algorithm) could not attain a generic quadratic speedup with respect to the classical hitting time \cite{shanleonew2019}. Moreover, it was not clear how to apply this algorithm to searching any ergodic, reversible Markov chain nor how it performs in general when there are multiple solutions. 

In this work we have proposed two new algorithms, based on CTQW, to find marked elements on any ergodic, reversible Markov chain. The first algorithm, which we refer to as the $\C\G'$ algorithm, is based on a modified version of the original $\C\G$ algorithm and uses a Hamiltonian that generates a quantum walk on the edges of the underlying Markov chain instead of its vertices. We obtain a general result regarding the performance of this algorithm and show it improves over the running time of the $\C\G$ algorithm for certain graphs (for example for low dimensional lattices), but fails to do so in general. However, it is not straightforward to relate the performance of this algorithm to the classical hitting time. Furthermore, the performance of this algorithm remains unclear when multiple vertices are marked.     

Our second algorithm, based on a randomized time-evolution of a Hamiltonian encoding an interpolated Markov chain,  surmounts several of these problems. Given any ergodic, reversible Markov chain $P$ with a set $M$ of marked elements, it can find an element of this marked set in a time that is equal to the square root of the extended hitting time of $P$ with respect to $M$. This implies a full quadratic speedup over its classical counterpart in the scenario where a single node is marked. 

Our results can lead to several new quantum algorithms. For example, it can be used to obtain analog quantum algorithms to prepare the stationary state of any ergodic, reversible Markov chain \cite{chakraborty2019analog}, a task that is used, for example by Google to rank webpages \cite{page1999pagerank}. To the best of our knowledge no analog quantum algorithms exist for quantum PageRank \cite{paparo2013quantum}. Our results could also lead to new quantum algorithms for quantum metropolis sampling \cite{temme2011quantum, yung2012quantum, ozols2013quantum}. 

Furthermore, our results use a novel approach for CTQW which is related to the corresponding DTQW framework. For example, the Hamiltonian implementing our CTQW-based algorithms can be simulated using query access to the unitary implementing the DTQW of Ref~\cite{krovi2016quantum} (See Appendix \ref{sec:quantum-simulation}). As a result, this opens up possibility of designing novel CTQW-based quantum algorithms for problems that have been tackled only in the DTQW framework.

For example, it has been recently observed that DTQW, in the framework of Ref.~\cite{krovi2016quantum}, can fast-forward the dynamics of any ergodic, reversible Markov chain \cite{simon2018quantum}. It would be natural to ask whether the same holds for CTQW as well. In fact, after the preparation of the first version of this manuscript, Ambainis et al. made use of this quantum fast forwarding algorithm to provide a quantum spatial search algorithm by DTQW, which even in the scenario where multiple nodes are marked, runs in square root of the classical hitting time (up to logarithmic overheads) \cite{ambainis2019quadratic}. It would be interesting to explore whether some of the techniques used therein can be used to obtain a full quadratic speedup in the CTQW framework.


\begin{acknowledgments}
S.C. and L.N. acknowledge funding from F.R.S.-FNRS. S.C. and J.R. are supported by the Belgian Fonds de la Recherche Scientifique - FNRS under grants no F.4515.16 (QUICTIME) and R.50.05.18.F (QuantAlgo). L.N. also acknowledges funding from Wiener-Anspach Foundation.
\end{acknowledgments}
\widetext
\clearpage
\begin{center}
\textbf{\large Appendix}
\end{center}
\setcounter{equation}{0}
\setcounter{figure}{0}
\setcounter{table}{0}
\setcounter{section}{0}
\setcounter{fact}{0}
\makeatletter
\renewcommand{\theequation}{S\arabic{equation}}
\renewcommand{\thefigure}{S\arabic{figure}}
\renewcommand{\thesection}{S\arabic{section}}
\renewcommand{\thetheorem}{S\arabic{theorem}}
\renewcommand{\thefact}{S\arabic{fact}}
\renewcommand{\thelemma}{S\arabic{lemma}}
\renewcommand{\thedefinition}{S\arabic{definition}}

\section{Proof of Theorem \ref{thm-main:search-somma-ortiz}}
\label{sec:proof-search-somma-ortiz}
Before proving Theorem \ref{thm-main:search-somma-ortiz}, we provide a general Lemma regarding quantum search with Hamiltonians fulfilling the following properties:
~\\
\begin{itemize}
\item[(i)]~ The Hamiltonian $H$ has dimension $2n-1$ and its eigenvalues are symmetric around $0$ i.e, $\lambda'_i=-\lambda'_{2n-i}$ for $i \in \{1, 2,...,2n-1\}$, implying that $\lambda'_n=0$. In addition, $H$ is normalized ($\nrm{H}=1$).~\\

\item[(ii)]~ For a particular quantum state $\ket{w}$, $H$ satisfies $\braket{w}{H|w}=0$. Furthermore, if we denote the eigenstates of $H$ as $\ket{v'_i}$ and the overlaps $a_i=\braket{w}{v'_i}$, we have that $|a_i|=|a_{2n-i}|$ for $i \in \{1, 2,...,2n-1\}$.
\end{itemize}
~\\
We demonstrate later that the Somma Ortiz Hamiltonian used in Algorithm~\ref{algo-cg-modified}
satisfies the aforementioned properties. These properties are also satisfied, for example, for certain lattice Hamiltonians \cite{foulger2014quantum, childs2014spatial-crystal}.  

Consider the search Hamiltonian
\begin{equation}
H_{\mathrm{search}}=H_{\mathrm{oracle}}+H=-\ket{w}\bra{w}H-H\ket{w}\bra{w}+H,
\end{equation} 
used in Algorithm \ref{algo-cg-modified}. We can show the following result.
~\\ 
\begin{lemma}
\label{lem:optimal-search-symmetric-spectrum}
Let $H$ be a Hamiltonian obeying the properties (i) and (ii) described above, and let us define the spectral gap $\Delta'=|\lambda_n-\lambda_{n-1}|=|\lambda_n-\lambda_{n+1}|$.  Furthermore, we define the overlap  $\sqrt{\epsilon}=|a_n|$ and the parameter
\begin{equation}\label{eq:mu_lemma}
\mu=\sqrt{\sum_{i\neq n}\dfrac{|a_i|^2}{\lambda^{'2}_i}}.
\end{equation}
Then provided there exists a small positive constant $c$ such that  $\sqrt{\epsilon}\leq c \Delta'\mu$, the evolution of the initial state $\ket{v'_n}$ under the Hamiltonian $H_{\text{search}}$ for time 
\begin{equation}\label{eq:time_lemma}
T=\frac{\pi}{2}\dfrac{\mu}{\sqrt{\epsilon}},    
\end{equation}
prepares a state $\ket{\tilde{f}}$ such that
\begin{equation}\label{eq:amp_lemma}
\nu =|\braket{\widetilde{w}}{\widetilde{f}}|= \Theta\left(\dfrac{1}{\mu\nrm{H\ket{w}}}\right),    
\end{equation}
where $\ket{\widetilde{w}}=H\ket{w}/\nrm{H\ket{w}}$ .
\end{lemma}
~\\
\begin{proof}
To demonstrate this, we compute the most relevant eigenvalues and eigenvectors of $H_{\text{search}}$ to obtain the  approximate dynamics. Note that, as $\braket{w}{H|w}=0$, we have that $H_{\text{search}}\ket{w}=0$. We want to find the conditions for which a vector $\ket{v}=\sum_i b_i\ket{v'_i}$ is an eigenstate of $H_{\text{search}}$ with non-zero eigenvalue $\lambda$, i.e.
\begin{equation}
\label{eq:condition_evalues-evectors}
H_{\text{search}}\ket{v}=\lambda\ket{v}.
\end{equation}
For any such $\ket{v}$, we have $\braket{w}{H_{\text{search}}|v}=\lambda\braket{w}{v}=0$, implying that $\braket{w}{v}=0$. Using this fact, we obtain from Eq.~\eqref{eq:condition_evalues-evectors} that 
\begin{equation}
\label{eq:b-i}
b_i=\dfrac{\gamma a_i}{\lambda'_i-\lambda},
\end{equation}  
where 
\begin{equation}
\label{eq:gamma}
\gamma=\braket{w}{H|v}=\sum_i\lambda'_i a_i^*b_i.
\end{equation}
Using Eq.~\eqref{eq:b-i} in Eq.~\eqref{eq:gamma}, we obtain that 
\begin{align}
\label{eq:eigenvalue_condition}
\sum_{i}\dfrac{|a_i|^2\lambda'_i}{\lambda'_i-\lambda}=1,
\end{align}
which gives the condition for $\lambda$ to be an eigenvalue. In fact, each interval $[\lambda'_{i},\lambda'_{i-1}]$ contains exactly one eigenvalue, since the LHS of this equation has poles at $\lambda=\lambda'_i$ and this function is monotonically decreasing within each of these intervals. We are interested in computing the eigenvalues that lie between $[\lambda'_{n-1},\lambda'_{n}]$ and $[\lambda'_{n},\lambda'_{n+1}]$. We denote these eigenvalues as $\delta_-$ and $\delta_+$, respectively, and its corresponding eigenstates as $\ket{v_+}$ and $\ket{v_-}$. Since $\lambda'_n=0$, we have that $\delta_+$ is is positive while $\delta_-$ is negative, with $|\delta^\pm|<\Delta'$.

We will now show that these solutions of  Eq.~\eqref{eq:eigenvalue_condition} lie in the following intervals
\begin{equation}
\label{eq:solutions-evalue-condition}
|\delta_\pm|\in \left[(1-\eta)\delta_0,(1+\eta)\delta_0\right],
\end{equation}
where
\begin{equation}
\label{eq:delta-0}
\delta_0=\dfrac{|a_n|}{\mu},
\end{equation}
and
\begin{equation}
\label{eq:eta}
\eta=\dfrac{|a_n|^2}{\mu^2\Delta'^2}.
\end{equation}
Note that the condition $\sqrt{\epsilon}=|a_n|\leq c \Delta' \mu$ ensures that $\eta\leq c^2$, where $c$ is a small constant.

In order to demonstrate this, we define the function
\begin{align}
F(\delta)&=\sum_{i}\dfrac{|a_i|^2\lambda'_i}{\lambda'_i-\delta}-1
\end{align}
and show that $F(\delta)$ is positive at $(1-\eta)\delta_0$ and $-(1+\eta)\delta_0$, and negative at $(1-\eta)\delta_0$ and $-(1-\eta)\delta_0$.

First, we note that $F(\delta)$ can be expanded as
\begin{align}
F(\delta)&=-|a_n|^2+\sum_{i\neq n}|a_i|^2\sum_{k=1}^\infty \left(\dfrac{\delta}{\lambda'_i}\right)^{k}\\
&=-|a_n|^2+\sum_{i\neq n}|a_i|^2\sum_{k=1}^\infty \left(\dfrac{\delta}{\lambda'_i}\right)^{2k}, \\
\end{align}
where in the second step we use the fact that terms of the form 
\begin{equation}\label{eq:symmetry_prop}
\sum_{i\neq n}\frac{|a_i|^2}{{\lambda'_i}^k}=0~~~\text{for odd}~k,     
\end{equation}
which follows from the symmetry properties of the spectrum (i) and (ii) which state that $\lambda'_i=-\lambda'_{2n-i}$ and $|a_i|=|a_{2n-i}|$, for $i\in \{1,2,...2n-1\}$.

We can now approximate $F(\delta)$ by noting that
\begin{align}
F(\delta)&=
|a_n|^2\left \{-1+\dfrac{1}{|a_n|^2}\sum_{i\neq n}\dfrac{|a_i|^2\delta^2}{\lambda^{'2}_i}+R(\delta)\right\},
\end{align}
with an error term
\begin{align}
R(\delta)&=\dfrac{1}{|a_n|^2}\sum_{i\neq n}\dfrac{|a_i|^2\delta^4}{\lambda^{'4}_i}\dfrac{1}{1-\frac{\delta^2}{\lambda^{'2}_i}}.
\end{align}
This quantity can be bounded for any value $\delta_{\pm}$ in the intervals from Eq.~\eqref{eq:solutions-evalue-condition} as 
\begin{align}
R(\delta_\pm)&\leq |a_n|^2\sum_{i\neq n}\dfrac{|a_i|^2}{\lambda^{'4}_i\mu^4}(1+O(c^2))\leq \dfrac{|a_n|^2}{\mu^2\Delta^2}(1+O(c^2))= \eta (1+O(c^2)).
\end{align}

Now we evaluate $F(\delta)$ at the points $\delta=(1\pm\eta)\delta_0$, obtaining
\begin{align}
F((1\pm\eta)\delta_0)&=|a_n|^2\left\{-1+(1\pm\eta)^2+R((1\pm\eta)\delta_0)\right\}\\
           &=|a_n|^2\left\{\pm 2\eta+\eta^2+R((1\pm\eta)\delta_0)\right\}.
\end{align}
Since $R((1\pm\eta)\delta_0)\leq \eta (1+O(c^2))$, we find that $F(\delta)$ is positive at $\delta=(1+\eta)\delta_0$ and negative at $\delta=(1-\eta)\delta_0$, for small enough $c$. Similar arguments also show that
$F(\delta)$ changes sign when evaluated at $-(1\pm\eta)\delta_0$.

Now that we have two approximate solutions of Eq.~\eqref{eq:eigenvalue_condition}, we proceed to estimating the value of $\gamma_{\pm}=\braket{w}{H |v_{\pm}}$ from Eq.~\eqref{eq:gamma}. Note that by substituting the value of $b_i$ from \eqref{eq:b-i} in the normalization condition $\sum_{i}|b_i|^2=1$ we have that 
\begin{align}
|\gamma_\pm|^2&=\left[\sum_{i}\dfrac{|a_i|^2}{(\delta_{\pm}-\lambda'_i)^2}\right]^{-1}\\
			  &=\left[\dfrac{|a_n|^2}{\delta^2_\pm}+\sum_{i\neq n}\dfrac{|a_i|^2}{\lambda_i^{'2}}\left(1-\frac{\delta_\pm}{\lambda'_i}\right)^{-2}\right]^{-1}\\
			  &=\left[\dfrac{|a_n|^2}{\delta^2_\pm}+\sum_{i\neq n}\dfrac{|a_i|^2}{{\lambda'_i}^2}+O\left(\sum_{i\neq n}\dfrac{|a_i|^2\delta^2_\pm}{{\lambda'_i}^4}\right)\right]^{-1}\\
			  &=\dfrac{1}{2\mu^2} \left(1+O(\eta)\right),
\end{align}
where in the second step we used again the symmetry property from Eq.~\eqref{eq:symmetry_prop}. Without loss of generality, we can choose $\gamma_{\pm}$ as well as $a_n$ to be positive, which allows us to estimate
\begin{equation}
b^\pm_n=\dfrac{\gamma_\pm a_n}{\delta_\pm}=\pm\dfrac{1}{\sqrt{2}}\left(1+\Theta(\eta)\right).
\end{equation}
Thus we have that the initial state, 
\begin{equation}
\ket{v'_n}= \dfrac{\ket{v_+}-\ket{v_-}}{\sqrt{2}}+\ket{\Phi},
\end{equation}
where $\nrm{\ket{\Phi}}=O(\eta)$. Then after a time $t=\frac{\pi}{2|\delta_\pm|}=\Theta(\frac{\mu}{\sqrt{\epsilon}})$,
\begin{equation}
e^{-iHt}\ket{v'_n}=\ket{\widetilde{f}},
\end{equation}
where 
\begin{equation}
\ket{\widetilde{f}}=\dfrac{\ket{v_+}+\ket{v_-}}{\sqrt{2}}+\ket{\Phi'},
\end{equation}
where $\nrm{\ket{\Phi'}}=O(\eta).$
Then observe that
\begin{align}
|\braket{\widetilde{w}}{f}| &= \dfrac{1}{\sqrt{2}\nrm{H\ket{w}}}\left(\gamma_++\gamma_-\right) + O(\eta)\\
                              &=\dfrac{1}{\mu\nrm{H\ket{w}}}+O(\eta)\\
                              &=\dfrac{1}{\mu\nrm{H\ket{w}}}\left(1+O\left(\dfrac{\epsilon}{\Delta'^2\mu}\right)\right)=\Theta\left(\dfrac{1}{\mu\nrm{H\ket{w}}}\right),
\end{align}
where in the last line we have used the condition that $\sqrt{\epsilon}\leq c\Delta'\sqrt{\mu}$ and that $||H\ket{w}||\leq 1$.
\end{proof}
\\~\\
We can now prove the following theorem from the main text, which we restate here.
\\~\\
\searchcgprime*
~\\
\begin{proof}
 First, we note that the analysis of Algorithm~\ref{algo-cg-modified} can be restricted to a $2n-1$ dimensional subspace of $\mathcal{H}\otimes\mathcal{H}$ given by $\mathcal{B}=\bigoplus_{k=1}^n\mathcal{B}_k$ (see Sec.~\ref{subsec:spectrum-somma-ortiz}). This results from the fact that the initial state $\ket{v_n,0}\in \mathcal{B}$ and that $\mathcal{B}$ is an invariant subspace of both $H=i[V^{\dag}SV,\Pi_0]$ and $H_{\mathrm{oracle}}$, as the solution state  $\ket{w,0}$ also belongs to this subspace.
 
 We can use Lemma~\ref{lem:optimal-search-symmetric-spectrum} to approximate the state obtained after step 2 of Algorithm \ref{algo-cg-modified}, as the Hamiltonian $H$ restricted to subspace $\mathcal{B}$ obeys the necessary properties (i) and (ii) required in the Lemma:
 ~\\
\begin{itemize}
\item[(i)]~ From Subsec.~\ref{subsec:spectrum-somma-ortiz}, we can see that the spectrum of the Somma-Ortiz Hamiltonian is symmetric around $0$, as the $0$-eigenstate of $H$ is $\ket{v_n,0}$ and the other eigenvalues are $E^{\pm}_k=\pm\sqrt{1-\lambda_k^2}$, for $k\in \{1,...,n-1\}$, with $\lambda_k$ being the eigenvalues of the discriminant matrix $D(P)$. ~\\
\item[(ii)]~It is easy to verify that $\braket{w,0}{H|w,0}=0$, since $\Pi_0 H \Pi_0=0$. Furthermore, $\braket{w,0}{\Psi^{\pm}_k}=\braket{w}{v_k}=a_k$, where $\ket{v_k}$ are eigenstates of $D(P)$, for $k\in \{1,...,n-1\}$ (see Eq.~\eqref{eqmain:spectrumSOham}).
\end{itemize}
~\\
Hence, the parameter $\mu$ defined in Lemma \ref{lem:optimal-search-symmetric-spectrum}, for the Somma-Ortiz Hamiltonian is given by 
\begin{equation}
\mu=\sqrt{\sum_{i=1}^{ n-1}\dfrac{2 |a_i|^2}{1-\lambda_i^2}}.    
\end{equation}
Moreover, for the Somma-Ortiz Hamiltonian the gap between the $0$-eigenvalue and its closest eigenvalues is $\Delta'=\Theta(\sqrt{\Delta})$, where $\Delta$ is the spectral gap of $D(P)$. Thus the validity condition in Lemma \ref{lem:optimal-search-symmetric-spectrum} becomes $\sqrt{\epsilon}\leq c\mu\sqrt{\Delta}$,  for some small constant $c$, as given by Eq.~\eqref{eqmain:spec_con}.
Provided this condition is satisfied, we can conclude from Lemma \ref{lem:optimal-search-symmetric-spectrum} that the evolution under $H_{\mathrm{search}}$ for time $T_1=\frac{\pi}{2}\frac{\mu}{\sqrt{\epsilon}}$ results in the state
\begin{equation}
\ket{\widetilde{f}}=\nu \ket{\widetilde{w}}+\sqrt{\epsilon}\ket{w,0}+\ket{\widetilde{w}}^\perp,
\end{equation}
where $\ket{\widetilde{w}}=H\ket{w,0}/||H\ket{w,0}$ and $\ket{\widetilde{w}}^\perp$ is an (unnormalized) quantum state such that orthogonal to both $\ket{w,0}$ and $\ket{\widetilde{w}}$. The amplitude $\nu$ is given by $\nu=\Theta(\mu^{-1} ||H\ket{w,0}||^{-1})$ from Eq.~\eqref{eq:amp_lemma}.

Step 3 of Algorithm~\ref{algo-cg-modified} applies the time evolution under $H_{\mathrm{oracle}}$ for time from Eq.~\eqref{eqmain:oracleCGprime} to this state. This Hamiltonian can be written as
\begin{equation}
 H_{\mathrm{oracle}}=- ||H\ket{w,0}|| \left( \ket{w,0}\bra{\widetilde{w}}+\ket{\widetilde{w}}\bra{w,0}\right).   
\end{equation}
As $\braket{w,0}{\widetilde{w}}=0$ this generates a rotation in a 2-d subspace spanned by $\ket{w,0}$ and $\ket{\widetilde{w}}$. Hence, the state after step 3 can be written (up to a global phase) as 
\begin{equation}
\ket{f}=\exp\left(-i\frac{\pi}{2} \frac{H_{\mathrm{oracle}}}{||H\ket{w,0}||}\right)\ket{\widetilde{f}}=\nu \ket{w,0}+\sqrt{\epsilon}\ket{\widetilde{w}}+\ket{\widetilde{w}}^\perp,
\end{equation}
which has an overlap $\nu$ with the marked node $\ket{w}$. As shown in the main article via Eqs.~\eqref{eqmain:bound1} and \eqref{eqmain:bound2} the evolution time $T_2=\Theta(\mu)$ and so the total evolution time $T=T_1+T_2=\Theta(\mu/\sqrt{\epsilon})$.
\end{proof}
\section{Worst-case performance of $\C\G'$ algorithm}
\label{sec:worstcaseCGprime}
In this section, we give an example of a Markov chain for which the parameter $\mu=\Theta(\Delta^{-1/2})$. For this example, Theorem~2 predicts that the $\C \G'$ algorithm for searching a marked node does not achieve a quadratic speed-up with respect to the hitting time of the corresponding classical walk. 

This example is based on a weighted Rook's graph, a graph whose connectivity represents the possible movements of a Rook on a rectangular chessboard of dimensions $n_1\times n_2$. From any position on the chessboard, the Rook can move vertically with probability $p$ and horizontally with probability $1-p$. If it moves vertically,  the probability of choosing any of the $n_1-1$ available positions is uniform. Similarly, if it moves vertically, it chooses any of $n_2-1$ with equal probability. The Markov chain corresponding to this random walk is 
\begin{equation}\label{eq:PweightedCG}
    P=\frac{p}{n_1-1}A^{n_1}_{CG}\otimes I_{n_2} +\frac{1-p}{n_2-1} I_{n_1}\otimes A^{n_2}_{CG},  
\end{equation}
where $A^{m}_{CG}$ denotes the adjacency matrix of the complete graph of $m$ nodes and $I_m$ the identity matrix of size $m$. Note that the unweighted walk on the Rook's graph considered in \cite{shanleonew2019} corresponds to $p=(n_1-1)/(n_1+n_2-2)$, in  which case the walker moves along any of the existing edges with equal probability. Note that $P$ is symmetric and so the discriminant matrix $D(P)=P$.

If $n$ is the total number of nodes, it can be seen that the eigenstate with eigenvalue $1$ is $\ket{s}=n^{-1/2}\sum_{i=1}^n \ket{i}$. The other eigenvalues are $p$ with degeneracy $n_2$, $1-p$ with degeneracy $n_1$ and $0$ with degeneracy $(n_1-1)(n_2-1)$. If $p=o(1)$, the spectral gap is given by $\Delta=p$. Furthermore, we can choose an orthogonal eigenbasis of the form 
\begin{equation}
\ket{\psi_j}=n^{-1/2}\sum_{i=1}^n\exp(i\phi_{k,j}\ket{j})    
\end{equation}
where $\phi_{k,j}$ are phases. The parameters $\mu$ and $||H\ket{w}||$, which determine the performance of the $\C\G'$ algorithm via the result in Theorem~\ref{thm-main:search-somma-ortiz}, are given by 
\begin{align}
    ||H\ket{w}||=\sqrt{2}\sqrt{\frac{n_1}{n}2p(1-p)+\frac{n_2}{n}(1-p^2)+\frac{(n_1-1)(n_2-1)}{n}}\\
     \mu=\sqrt{2}\sqrt{\frac{n_1}{n}\frac{1}{2p(1-p)}+\frac{n_2}{n}\frac{1}{(1-p^2)}+\frac{(n_1-1)(n_2-1)}{n}}
\end{align}
Let us choose $n_1=n/\ell$ and $n_2=\ell$ for some fixed positive integer $\ell=\Theta(1)$ and $p$ such that $\Delta=p=o(1)$. For this choice, we have that $||H\ket{w}||=\Theta(1)$ and 
$$\mu=\Theta\left(\dfrac{1}{\sqrt{\Delta}}\right)=\Theta\left(\dfrac{1}{\sqrt{p}}\right)$$. 

We also choose $p$ large enough so that spectral condition that is necessary for Theorem~\ref{thm-main:search-somma-ortiz} to hold. This is valid if $p\gg 1/n$. Given this choice, the maximum amplitude reached at the marked node via the $\C\G'$ algorithm is
$$
\nu=\Theta(\sqrt{\Delta})=\Theta(\sqrt{p}),
$$ 
in time 
$$
T=\Theta\left(\dfrac{1}{\sqrt{\epsilon\Delta}}\right)=\Theta\left(\sqrt{\dfrac{n}{p}}\right).
$$ 
Hence, this fails to achieve a quadratic speed-up with respect to the classical hitting time which is upper bounded as
$$HT(P,w)\leq\Oo\left(\dfrac{1}{\epsilon\Delta}\right)=\Oo\left(\dfrac{n}{p}\right).$$

Interestingly, for this choice of $p$ and $n_1$, the original Childs and Goldstone approach outperforms the $\C\G'$ algorithm and, in fact, runs in optimal time. This can be derived from the necessary and sufficient conditions for optimality of this algorithm, presented in Ref.~\cite{shanleonew2019}. Therein, we show that for any graph with spectral gap $\Delta\gg\epsilon$, the $\C\G$ algorithm to be optimal is iff
\begin{equation}\label{eq:optimalityconCG}
    \frac{S_1}{\sqrt{S_2}}=\Theta(1), 
\end{equation}
where 
\begin{equation}
    S_k=\sum_{i=1}^{n-1}\frac{|a_i|^2}{(1-\lambda_i)^k},
\end{equation}
for positive integer $k$. It can be seen that the parameters $S_1$ and $S_2$ scale as $S_1=\Theta(\Delta^{-1})$ and $S_2=\Theta(\Delta^{-2})$, implying that the optimality condition of Eq.~\eqref{eq:optimalityconCG} is obeyed. Thus, while the $\C\G'$ algorithm improves upon $\C\G$ in some important examples, it fails to do so in general.   
\section{A brief overview on quantum phase randomization}
\label{sec:quantum-phase-randomization}
We shall now briefly discuss the technique of quantum phase randomization introduced by Boixo et al. \cite{boixo2009eigenpath}. The main idea is that one can approximate \textit{idealized} projective measurements by randomized evolutions. 

Consider a Hamiltonian $H$ with eigenvalues, $\lambda_1 \geq \lambda_2\geq \cdots\geq \lambda_{n-1}> \lambda_n=0$ and the corresponding eigenvectors, $\ket{\lambda_1},\cdots,\ket{\lambda_n}$, respectively. Let $\rho_0=\ket{\psi_0}\bra{\psi_0}$, where $\ket{\psi_0}=\sum_{k=1}^{n}c_k\ket{\lambda_k}$. Also let $\mathcal{U}^t(\rho)$ be the quantum operation corresponding to evolving a state $\rho$ under $H$ for a time $t$. Applying this operation to the state $\rho_0$, we obtain the following quantum state
\begin{align}
\label{eq:time_evolution_state}
\mathcal{U}^t(\rho_0)=\ket{\psi(t)}\bra{\psi(t)}&=e^{-iHt}\ket{\psi_0}\bra{\psi_0}e^{iHt}\\
                          &=|c_n|^2\ket{\lambda_n}\bra{\lambda_n}+\sum_{k=1}^{n} e^{i\lambda_k t} c_n c_k^*\ket{\lambda_n}\bra{\lambda_k} + \sum_{k,l\neq n} e^{-it(\lambda_k-\lambda_l)}c_k c_l^*\ket{\lambda_k}\bra{\lambda_l}+h.c. 
\end{align}

Now consider an \textit{idealized} measurement process with POVM $\{\ket{\lambda_n}\bra{\lambda_n},~I-\ket{\lambda_n}\bra{\lambda_n}\}$, followed by an operation $\E$, that has no effect on $\ket{\lambda_n}$ (acts on the space orthogonal to $\ket{\lambda_n}$). That is,
\begin{align}
\label{eq:measurement_general}
\mathcal{M}^{\E}_n(\rho)&=\ket{\lambda_n}\bra{\lambda_n}\rho_0\ket{\lambda_n}\bra{\lambda_n}+\E\left[(I-\ket{\lambda_n}\bra{\lambda_n})\rho_0(I-\ket{\lambda_n}\bra{\lambda_n})\right].
\end{align}
If $\E$ is the time evolution operation, i.e.\ $\E=\U^t$, then we obtain that
\begin{equation}
\label{eq:measurement_time_evolution}
\mathcal{M}^{\U^t}_n(\rho)=|c_n|^2\ket{\lambda_n}\bra{\lambda_n}+\sum_{k,l\neq n} e^{-i(\lambda_k-\lambda_l)t}c_k c_l^*\ket{\lambda_k}\bra{\lambda_l}+h.c.
\end{equation}
Clearly, one obtains the eigenstate $\ket{\lambda_n}$ with probability $|c_n|^2$. 
Observe that
\begin{equation}
\label{eq:approx_measurement_evolution}
\nrm{\U^t(\rho_o)-\M^{\U^t}_n(\rho_o)}=\nrm{\sum_k e^{i\lambda_k t}c_n c^*_k \ket{\lambda_n}\bra{\lambda_k}}+h.c.,
\end{equation}
i.e.\ they differ only in the coherences. In what follows, we show how to bound these coherences, i.e. the RHS of Eq.~\eqref{eq:approx_measurement_evolution} by a small constant $\eps$ in order to approximate the idealized operation $\M^{\U}_n(\rho_0)$ by $\U^t(\rho_0)$ up to an error $\eps$. 

To achieve this, we will consider that the time of evolution $t$ is a random variable from some probability distribution $\mu$. Such a randomized time evolution introduces dephasing in the eigenbasis of the Hamiltonian. In such a scenario, Boixo et al. show that is possible to bound the coherences in terms of the characteristic function of the underlying distribution. When $t$ is a random variable, we have that the randomized time evolution
\begin{equation}
\mathcal{\overline{U}}(\rho_0)=\int \mathcal{U}^t(\rho_0) d\mu,
\end{equation}
with $\mu$ being the probability distribution of $t$.
Then,
\begin{align}
\label{eq:bound_coherence}
\nrm{\overline{\U}(\rho_0)-\M^{\overline{\U}}_n(\rho_0)}&=\nrm{\sum_k \int e^{i\lambda_k t}c_n c^*_k \ket{\lambda_n}\bra{\lambda_k} d\mu+h.c.}\\
                                    &=\nrm{\sum_k c_n c^*_k \Phi(\lambda_k) \ket{\lambda_n}\bra{\lambda_k} +h.c.},
\end{align} 
where $\Phi(\omega)=\int e^{i\omega t}~d\mu$ is the characteristic function of the random variable $t$. If we consider the norm to be the Frobenius norm, we have
\begin{equation}
\nrm{\overline{\U}(\rho_0)-\M^{\overline{\U}}_n(\rho_0)}_F=\sqrt{2\sum_k \left|c_n c^*_k \Phi(\lambda_k)\right|^2}.
\end{equation}
In the following section we will show that by choosing a uniform distribution in a large enough time interval this term can be bounded. The average cost of randomized time evolution is $\langle t \rangle$.
\section{Proof of Theorem~\ref{thm-main:search-phase-randomization}}
\label{sec:proof-search-phase-randomization}
\begin{proof}
 The time-evolution of the state $\rho_0=\ket{v_n(0),0}\bra{v_n(0),0}$ under $H(s)$ for time $t$ is given by $\mathcal{U}^t(\rho_0)=\ket{\psi(t)}\bra{\psi(t)}$, where
\begin{equation}\label{eq:psit}
    \ket{\psi(t)}=\alpha_n \ket{v_n,0}+ \sum_{j=1}^{n-1}\sum_{\sigma=\pm} \alpha_j e^{-i t E_j^\sigma} \ket{\Psi_j^\sigma(s)}. 
\end{equation}
 The energies $E_j^\sigma$ are the eigenvalues of $H(s)$ in the subspace $\bigoplus_{k=1}^{n-1}\mathcal{B}_k$ (see Sec.~\ref{sec:SOformalism}), $\ket{\Psi_k^\sigma(s)}$ are its corresponding eigenstates and 
 \begin{align}
\label{eq:overlap-time-evolution}
\alpha_n&=\braket{v_n(s),0}{v_n(0),0}=\sqrt{1-p_M}\cos\theta(s)+\sqrt{p_M}\sin\theta(s),\\ \alpha_j&=\braket{\Psi_k^\sigma(s)}{v_n(0),0}=\dfrac{\sqrt{1-p_M}\braket{v_j(s)}{U}+ \sqrt{p_M}\braket{v_j(s)}{M}}{\sqrt{2}}. \label{eq:overlap2-time-evolution}   
\end{align}
where $\cos\theta(s)$ and $\sin\theta(s)$ are defined in Eqs.~\eqref{eqmain:costheta} and~\eqref{eqmain:sintheta}.

Let us consider the operations defined in Eqs.~\eqref{eq:time_evolution_state} and \eqref{eq:measurement_time_evolution}, which in our case are given by $\mathcal{U}^t(\rho_0)=\ket{\psi(t)}\bra{\psi(t)}$ and $\mathcal{M}_n^{\mathcal{U}^t}(\rho_0)$, where the latter results from a projective measurement on $\rho_0$ with measurement operators $\{\ket{v_n(s)}\bra{v_n(s)}, I-\ket{v_n(s)}\bra{v_n(s)}\}$ followed by time evolution for time $t$ with Hamiltonian $H(s)$. The state $\mathcal{M}_n^{\mathcal{U}^t}(\rho_0)$ has the form 
\begin{equation}\label{eq:operationM}
    \mathcal{M}_n^{\mathcal{U}^t}(\rho_0)=|\alpha_n^2|\ket{v_n(s),0}\bra{v_n(s),0}+ (1-|\alpha_n^2|)\rho'(t), 
\end{equation}
where $\rho'(t)$ is a quantum state in the subspace orthogonal to $\ket{v_n(s),0}$.  

Let us consider that $t$ is a random value chosen uniformly in the interval $[0,T]$, and denote the expected state after this randomized time-evolution as $\overline{\U^T}(\rho_0)$. Using Eqs. \eqref{eq:bound_coherence} and \eqref{eq:psit} we obtain that  
\begin{align}
\nrm{\overline{\U^T}(\rho_0)-\M^{\overline{\U^T}}_n(\rho_0)}&=\nrm{\dfrac{1}{T}\sum_{j=1}^{n-1}\sum_{\sigma=\pm}\int_0^T  dt~\alpha_j^* \alpha_n e^{it E_j^\sigma}\ket{v_n(s),0}\bra{\Psi^{\sigma}_j(s)}+h.c.}\\
                                   &=\nrm{\sum_{j=1}^{n-1} \dfrac{\alpha_j^* \alpha_n}{T}\left(\int_{0}^T dt e^{iE_j^+ t} \ket{v_n(s),0}\bra{\Psi^+_j(s)}+\int_{0}^T dt e^{iE_j^- t} \ket{v_n(s),0}\bra{\Psi^-_j(s)}\right)+h.c.}
                                  \end{align}
Let us define $\phi_j=|E_j^{\pm}|=\sqrt{1-\lambda_j(s)^2}$. This way, we can write
                                 \begin{align}                                          \nrm{\overline{\U^T}(\rho_0)-\M^{\overline{\U^T}}_n(\rho_0)}_F &=\nrm{\sum_{j=1}^{n-1} \dfrac{\alpha_j^* \alpha_n}{\phi_j T}\left((1-e^{iT\phi_j}) \ket{v_n(s),0}\bra{\Psi^+_j(s)}+(1-e^{-iT\phi_j}) \ket{v_n(s),0}\bra{\Psi^-_j(s)}\right)}_F\\
                                  &=2\sqrt{2\sum_{j=1}^{n-1} \dfrac{|\alpha_j^* \alpha_n|^2}{\phi^2_j T^2}\sin^2(\phi_j T/2)}\\
                                  &\leq 2\sqrt{\dfrac{2}{T^2}\sum_{j=1}^{n-1} \dfrac{|\alpha_j^* \alpha_n|^2}{1-\lambda_j(s)^2}}.
\end{align}
From now on, we choose
\begin{equation}
s=s^*=1-\dfrac{p_M}{1-p_M}    
\end{equation}
which ensures that $\ket{v_n(s^*)}$ has a large overlap with both the marked subspace as well as with the initial state  $\ket{v_n(0)}$ as discussed in Sec.~\ref{subsec:algorandomt}. For this choice, we have $\cos(s^*)=\sin(s^*)=1/\sqrt{2}$ (see Eq.~Eqs.~\eqref{eqmain:costheta} and~\eqref{eqmain:sintheta}) implying that 
\begin{equation}\label{eq:vns_star}
    \ket{v_n(s^*)}=\dfrac{\ket{U}+\ket{M}}{\sqrt{2}}.
\end{equation}
Moreover, from Eq.~\eqref{eq:overlap-time-evolution} we have that  
\begin{equation}\label{eq:alpha_n}
|\alpha_n|^2=\dfrac{1}{2}+\sqrt{p_M(1-p_M)}.    
\end{equation}
Furthermore, for any $1\leq j \leq n-1$, we have $\braket{v_j(s^*)}{v_n(s^*)}=0$ and so $\braket{v_j(s^*)}{U}=-\braket{v_n(s^*)}{M}$. Combining this with Eq.~\eqref{eq:overlap2-time-evolution} we have that when $s=s^*$,
\begin{equation}
|\alpha_j|^2=\left(\dfrac{1}{2}-\sqrt{2p_M(1-p_M)}\right)|\braket{v_j(s^*)}{U}|^2 \leq \left(\dfrac{1}{2}-\sqrt{p_M(1-p_M)}\right)|\braket{v_j(s^*)}{U}|^2.    
\end{equation}
Thus,
\begin{align}
\nrm{\overline{\U^T}(\rho_0)-\M^{\overline{\U^T}}_n(\rho_0)}&\leq \sqrt{\dfrac{\left(1/4-p_M(1-p_M)\right)}{T^2}\sum_j\dfrac{8|\braket{v_j(s^*)}{U}|^2}{1-\lambda_j(s^*)^2}}\\
                                  &\leq \sqrt{\dfrac{\left(1-4p_M(1-p_M)\right)}{T^2}\sum_j\dfrac{2|\braket{v_j(s^*)}{U}|^2}{1-\lambda_j(s^*)}}.
\end{align}
It is fair to assume that $p_M<1/4$, otherwise, one could simply prepare the state $\ket{v_n(0)}$ and measure, thereby obtaining, with a high probability, a marked vertex. Assuming this and by choosing 
\begin{equation}
T\geq  \dfrac{\sqrt{2HT(s^*)}}{\eps}=\dfrac{1}{\eps}\sqrt{\sum_{j=1}^{n-1}\dfrac{2|\braket{v_j(s^*)}{U}|^2}{1-\lambda_j(s^*)}}    
\end{equation} we obtain the bound
\begin{align}\label{eq:finalbound}
\nrm{\overline{\U^T}(\rho_0)-M^{\overline{\U^T}}_n(\rho_0)}&\leq \eps.
\end{align}
Note that from Eq.~\eqref{eqmain:interpolated-vs-extended-hitting-time}, $HT(s^*)=HT^+(P,M)/4$ and thus we have that
\begin{equation}
T\geq \dfrac{1}{\eps}\sqrt{\dfrac{HT^+(P,M)}{2}}.    
\end{equation}

Finally, from Eqs.~\eqref{eq:operationM}, \eqref{eq:vns_star} and \eqref{eq:alpha_n} we can see that 
\begin{equation}
    \mathrm{Tr}\left(\left(\Pi_M \otimes I\right)M^{\overline{\U^T}}_n(\rho_0)\right)\geq |\alpha_n|^2|\braket{v_n(s^*)}{M}|^2\geq \frac{1}{4}.
\end{equation}
Hence, if we define $\overline{\rho}(T)=\overline{\U^T}(\rho_0)$ and use the bound from Eq.~\eqref{eq:finalbound}, we conclude that the expected value of the probability of observing a marked vertex (success probability) after the final step of Algorithm~\ref{algo1} is lower bounded as 
\begin{align}
p_{\mathrm{succ}}&=\mathrm{Tr}\left(\left(\Pi_M \otimes I\right)\overline{\rho}(T)\right)\\
                 &\geq \mathrm{Tr}\left(\left(\Pi_M \otimes I\right)M^{\overline{\U^T}}_n(\rho_0)\right)-\eps\\
                 &\geq 1/4-\eps.
\end{align}
\end{proof}

\section{Cost of simulating the Somma-Ortiz Hamiltonian}
\label{sec:quantum-simulation}
In order to directly use recent quantum simulation algorithms to obtain our results, we shall make use of the \textit{block-encoding} framework \cite{chakraborty2018power}.
~\\
	\begin{definition}[Block-encoding \cite{low2016hamiltonian,chakraborty2018power}]\label{def:standardForm}
		Suppose that $A$ is an $s$-qubit operator, $\alpha,\eps\in\R_+$ and $a\in \mathbb{N}$. Then we say that the $(s+a)$-qubit unitary $U$ is an $(\alpha,a,\eps)$-block-encoding of $A$, if 
		$$ \nrm{A - \alpha(\bra{0}^{\otimes a}\otimes I)U(\ket{0}^{\otimes a}\otimes I)}\leq \eps. $$
	\end{definition}
~\\
Note that trivially, any unitary $U$ is a (1,0,0)-block encoding of itself. 
We establish the cost of simulating the Somma-Ortiz Hamiltonian $H(s)$ in terms of the number of queries made to the discrete-time quantum walk unitary for the spatial search problem of Ref.~\cite{krovi2016quantum}. This unitary is given by
\begin{equation}
    \label{eq:szegedy-interpolated-unitary}
    W(s)=V(s)\ S\ V(s)^\dag (2\Pi_0-I).
\end{equation}
Note that the Hamiltonian $H(s)$ can be written as
\begin{equation}
    \label{eq:ham-diff-unitaries}
H(s)=\dfrac{i}{2}\left(W(s)-W(s)^{\dagger}\right).     
\end{equation} 
By defining $\overline{W}(s)=iW(s)$, we can rewrite this as
\begin{equation}
    \label{eq:ham-sum-unitaries}
H(s)=\dfrac{\overline{W}(s)+\overline{W}(s)^{\dagger}}{2}.   
\end{equation}

Let us define the controlled unitary
\begin{equation}
    \widetilde{W}(s)=\ket{0}\bra{0}\otimes \overline{W}(s) + \ket{1}\bra{1}\otimes \overline{W}(s)^\dag
\end{equation}

Then we can define the following lemma:
\\~\\
\begin{lemma}
\label{lem:block-soma-ortiz}
If $Q$ is the single qubit Hadamard unitary and 
$$
U(s)=(Q\otimes I)\widetilde{W}(s)(Q\otimes I),
$$
then $U(s)$ is a (1,1,0)-block encoding of $H(s)$.
\end{lemma}
~\\
\begin{proof}
 We have that 
 \begin{align}
     U(s)\ket{0}\ket{\psi}&=(Q\otimes I)\widetilde{W}(s)(Q\otimes I)\ket{0}\ket{\psi}\\
                       &=(Q\otimes I)\widetilde{W}(s)\ket{+}\ket{\psi}\\
                       &=\dfrac{1}{\sqrt{2}}\left(Q\otimes I\right)\left(\ket{0}\overline{W}(s)\ket{\psi}+\ket{1}\overline{W}(s)^\dag\ket{\psi}\right)\\
                       &=\dfrac{1}{\sqrt{2}}\left(\ket{+}\overline{W}(s)\ket{\psi}+\ket{-}\overline{W}(s)^\dag\ket{\psi}\right)\\
                       &=\ket{0}H(s)\ket{\psi}+\ket{\Phi},
 \end{align}
 where $\ket{\Phi}$ is an unnormalized state such that $(\ket{0}\bra{0}\otimes I)\ket{\Phi}=0$.
\end{proof}

Now we are in a position to directly use the results of Low and Chuang \cite{low2016hamiltonian}. We first state their general result formally
\begin{theorem}[Hamiltonian simulation of block-encoded matrices \cite{low2016hamiltonian}]
\label{thm:blockHamSim}
		Suppose that $U$ is an $(\alpha,a,0)$-block-encoding of the Hamiltonian $H$. Then we can implement an $\eps$-precise Hamiltonian simulation unitary $B$ which is an $(1,a+2,\eps)$-block-encoding of $e^{-itH}$, with $\bigO{|\alpha t|+\log(1/\eps)}$ uses of $U$ .
\end{theorem}
~\\
We now apply Theorem \ref{thm:blockHamSim} to the result obtained from Lemma \ref{lem:block-soma-ortiz}. An $\eps$-precise quantum simulation of $H(s)$, i.e.\ an $(1,3,\eps)$-block encoding of $e^{-itH(s)}$ can be implemented using $\Oo(t+\log (1/\eps))$-calls to $U(s)$. As $U(s)$ can be constructed by only two calls to $W(s)$, we obtain the following fact:
~\\
\begin{fact}
\label{fact:simulation}
An $(1,3,\eps)$-block encoding of $e^{-itH(s)}$ can be implemented by using $\Oo(t+\log(1/\eps))$-queries to $W(s)$.
\end{fact}
~\\
This establishes a relationship between the CTQW-framework defined here and its discrete-time counterpart developed in Refs.~\cite{szegedy2004quantum,krovi2016quantum,magniez2011search}.

One can also state the cost of simulating $H(s)$ in terms of basic Markov chain operations which would enable us to express the complexity of our quantum spatial search algorithm (Algorithm \ref{algo1}) in terms to these operations. To that end, given a Markov chain $P$, let us define the following oracular operations:
~\\
\begin{itemize}
    \item Check ($M$): Cost of checking whether a given node is marked. We denote this by $\C$.
    \item Update ($P$): Cost of applying one step of the walk $P$, which we denote by $\U$.
    \item Setup ($P$): The cost of preparing the initial state $\ket{v_n}$ (in Eq.~\eqref{eqmain:initial-stationary-state}), denoted by $\mathcal{S}$.
\end{itemize}
~\\
From Refs.~\cite{krovi2016quantum,ambainis2019quadratic}, we know that the cost of implementing $W(s)$ is in $\Oo(\C+\U)$. As a result from Fact \ref{fact:simulation}, the running time of Algorithm \ref{algo1} is
\begin{equation}
    T=\Oo\left(\mathcal{S}+\sqrt{HT^+(P,M)}\left(\C+\U\right)\right).
\end{equation}


\bibliographystyle{unsrt}
\bibliography{References}
\end{document}